\documentclass[final,twocolumn,rmp,floats,amsfonts,amssymb,amsmath,superscriptaddress,showpacs,floatfix,a4paper]{revtex4}
\usepackage{graphicx}
\usepackage{natbib}
\newcommand{\degre}{$^{\rm o}$}
\newcommand{\zob}{${\cal{R}}_{\pi}$}
\newcommand{\ie}{\emph{i.e.}, }
\newcommand{\vs}{\emph{vs.} }

\begin{document}
\title{Supercritical transition to turbulence in an inertially-driven von K\'arm\'an closed flow.}

\author{Florent Ravelet}
\affiliation{Service de Physique de l'Etat Condens\'e, Direction des Sciences de la Mati\`ere, CEA-Saclay, CNRS URA 2464, 91191 Gif-sur-Yvette cedex, France}
\affiliation{Present address: Laboratory for Aero and Hydrodynamics, Leeghwaterstraat 21, 2628 CA Delft, The Netherlands}
\email{florent.ravelet@ensta.org}
\author{Arnaud Chiffaudel}
\affiliation{Service de Physique de l'Etat Condens\'e, Direction des Sciences de la Mati\`ere, CEA-Saclay, CNRS URA 2464, 91191 Gif-sur-Yvette cedex, France}
\author{Fran\c{c}ois Daviaud}
\affiliation{Service de Physique de l'Etat Condens\'e, Direction des Sciences de la Mati\`ere, CEA-Saclay, CNRS URA 2464, 91191 Gif-sur-Yvette cedex, France}

\date{accepted in JFM, January 18, 2008}


\begin{abstract}
We study the transition from laminar flow to fully developed turbulence for an inertially-driven von K\'arm\'an flow between two counter-rotating large impellers fitted with curved blades over a wide range of Reynolds number ($10^2-10^6$). The transition is driven by the destabilisation of the azimuthal shear-layer, \ie Kelvin--Helmholtz instability which exhibits travelling/drifting waves, modulated travelling waves and chaos below the emergence of a turbulent spectrum. A local quantity ---the energy of the velocity fluctuations at a given point--- and a global quantity ---the applied torque--- are used to monitor the dynamics. The local quantity defines a critical Reynolds number $Re_c$ for the onset of time-dependence in the flow, and an upper threshold/crossover $Re_t$ for the saturation of the energy cascade. The dimensionless drag coefficient, \ie the turbulent dissipation, reaches a plateau above this finite $Re_t$, as expected for a \lq\lq Kolmogorov\rq\rq-like turbulence for $Re \rightarrow \infty$. Our observations suggest that the transition to turbulence in this closed flow is globally supercritical: the energy of the velocity fluctuations can be considered as an order parameter characterizing the dynamics from the first laminar time-dependence up to the fully developed turbulence. Spectral analysis in temporal domain moreover reveals that almost all of the fluctuations energy is stored in time-scales one or two orders of magnitude slower than the time-scale based on impeller frequency.

\end{abstract}

\maketitle
\section{\label{sec:intro2}Introduction}
Hydrodynamic turbulence is a key feature for many common problems \cite[]{tennekes,lesieur}. In a few ideal cases, exact solutions of the Navier--Stokes equations are available, based on several assumptions such as auto-similarity, stationariness, or symmetry \cite[for a collection of examples, see][]{schlichting}. Unfortunately, they are often irrelevant in practice, because they are unstable. Two of the simplest examples are the centrifugal instability of the Taylor--Couette flow between two concentric cylinders, and the Rayleigh--B\'enard convection between two differentially heated plates: once the amount of angular momentum or heat is too important to be carried by molecular diffusion, a more efficient convective transport arises. Increasing further the control parameter in these two examples, secondary bifurcations occur, leading rapidly to temporal chaos, and/or to spatio-temporal chaos, then to turbulence.

Several approaches have been carried in parallel concerning developed turbulence, focused on statistical properties of flow quantities at small scales \cite[]{frisch95} or taking into account the persistence of coherent structures in a more deterministic point of view \cite[]{tennekes,lesieur}. One of the major difficulty concerning a self-consistent statistical treatment of turbulence is indeed that it depends on the flow in which it takes place (for instance wakes, jets or closed flows). At finite $Re$, most turbulent flows could still keep in average some geometrical or topological properties of the laminar flow (for example the presence of a B\'enard--von K\'arm\'an street in the wake of a bluff body whatever $Re$), which could still influence its statistical properties \cite[]{zocchi94,laporta01,ouellette2006}.  

Furthermore, we have recently shown in a von K\'arm\'an flow that a turbulent flow can exhibit multistability, first order bifurcations and can even keep traces of its history at very high Reynolds number \cite[]{ravelet2004}. The observation of this turbulent bifurcation lead us to study the transition from the laminar state to turbulence in this inertially driven closed flow. 

\subsection{\label{subsec:vkflow} Overview of the von K\'arm\'an swirling flow}
\subsubsection{Instabilities of the von K\'arm\'an swirling flow between flat disks}

The disk flow is an example where exact Navier--Stokes solutions are available. The original problem of the flow of a viscous fluid over an infinite rotating flat disk has been considered by \citet{vonkarman21}. Experimentally, the problem of an infinite disk in an infinite medium is difficult to address. Addition of a second coaxial disk has been proposed by \citet{batchelor51} and \citet{stewartson53}. A cylindrical housing to the flow can also be added. Instabilities and transitions have been extensively studied in this system \cite[for instance in][]{mellor68,harriott84,escudier84,sorensen95,gelfgat1996,spohn98,gauthier99,schouveiler01,nore03,nore04jfm,nore05}.  The basic principle of this flow is the following: a layer of fluid is carried near the disk by viscous friction and is thrown outwards by centrifugation. By incompressibility of the flow, fluid is pumped toward the centre of the disk. Since the review of \citet{zandbergen87}, this family of flow is called \lq\lq von K\'arm\'an swirling flow\rq\rq.  In all cases, it deals with the flow between smooth disks, at low-Reynolds numbers, enclosed or not into a cylindrical container.

\subsubsection{The \lq\lq French washing machine\rq\rq: an inertially-driven, highly turbulent von K\'arm\'an swirling flow.}

Experimentally, the so-called \lq\lq French washing-machine\rq\rq~ has been a basis for extensive studies of very high-Reynolds number turbulence in the last decade \cite[]{douady91,fauve93,zocchi94,cadot95,labbe96jphys,tabeling96,cadot97,laporta01,moisy01,bourgoin02,titon03pof,leprovost04,marie04pof,ravelet2004}. To reach a Kolmogorov regime in these studies, a von K\'arm\'an flow is inertially-driven between two disks fitted with blades, at a very high-Reynolds number ($10^5 \lesssim Re \lesssim 10^7$). Due to the inertial stirring, very high turbulence levels can be reached, with fluctuations up to $50 \%$ of the blades velocity, as we shall see in this article. 

Most of the inertially-driven von K\'arm\'an setups studied in the past dealt with straight blades. Von K\'arm\'an flows with curved-blades-impellers were first designed by the VKS-team for dynamo action in liquid sodium \cite[]{bourgoin02,monchaux2006b}. With curved blades, the directions of rotation are no longer equivalent. One sign of the curvature ---\ie with the convex face of the blades forward, direction~$(+)$--- has been shown to be the most favourable to dynamo action \cite[]{marie2003epjb,ravelet2005,monchaux2006b}. The turbulent bifurcation \cite[]{ravelet2004} has been obtained with the concave face of the blades forward, direction~$(-)$. In this last work, we discussed about the respective role of the turbulent fluctuations and of the changes in the mean-flow with increasing the Reynolds number on the multistability.

\subsection{\label{subsec:plan} Outline of the present article}

Our initial motivation to the present study was thus to get an overview of the transition to turbulence and to check the range where multistability exists. We first describe the experimental setup, the fluid properties and the measurement techniques in~\S\,\ref{sec:exp}. The main data presented in this article are gathered by driving our experiment continuously from laminar to turbulent regimes for this peculiar direction of rotation, covering a wide range of Reynolds numbers. In~\S\,\ref{sec:descr_regimes} and \S\,\ref{sec:quant} we characterise the basic flow and describe the transition from the laminar regime to turbulence through quasi-periodicity and chaos and explore the construction of the temporal spectrum of velocity fluctuations. The continuity and global supercriticality of the transition to turbulence is a main result of this article. 

In \S\,\ref{sec:visc2inert} we obtain complementary data by comparing the two different directions of rotation and the case with smooth disk. We show how inertial effects clearly discriminate both directions of rotation at high-Reynolds numbers.

We then summarize and discuss the main results in~\S\,\ref{sec:discussion}.

\section{Experimental setup} \label{sec:exp}

\begin{figure}
\begin{center}
\includegraphics[clip,width=45mm]{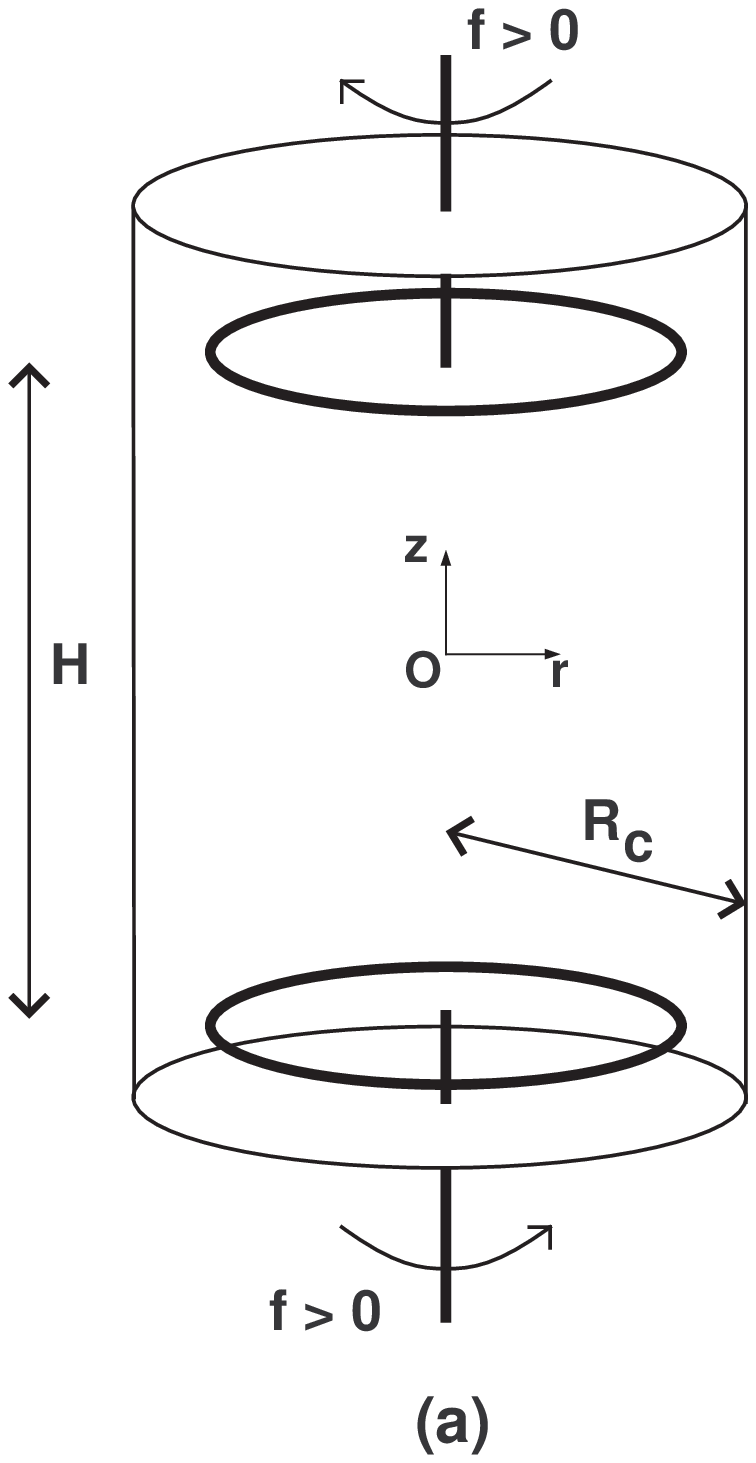}
\includegraphics[clip,width=40mm]{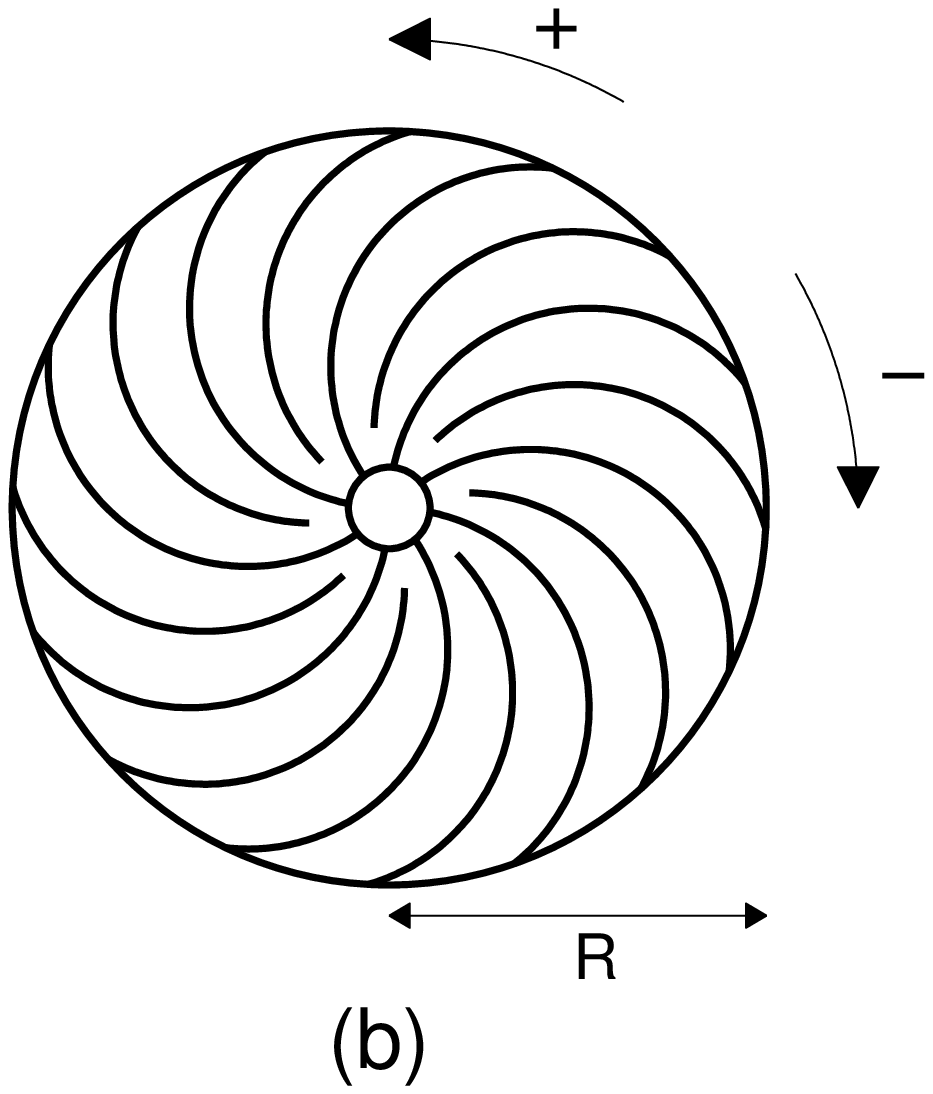}
\caption{(a) Sketch of the experiment. The flow volume between the impellers is of height $H=1.8 \ R_c$. (b) Impellers used in this article. The disks radius is $R=0.925 \ R_c$ and they are fitted with 16 curved blades: the two different directions of rotation defined here are not equivalent. This model of impellers has been used in the VKS1 sodium experiment \cite[]{bourgoin02} and is called TM60.}
\label{fig:sketch}
\end{center}
\end{figure}

\subsection{\label{subsec:dimension}Dimensions, symmetries and control parameter}
The cylinder radius and height are, respectively, $R_c=100 \ {\rm mm}$ and $H_c=500 \ {\rm mm}$. A sketch of the experiment can be found in figure~\ref{fig:sketch}(a). We use bladed disks to ensure inertial stirring. The impellers consist of $185 \ {\rm mm}$ diameter stainless-steel disks each fitted with 16 curved blades, of curvature radius $50 \ {\rm mm}$ and of height $h=20 \ {\rm mm}$ (figure~\ref{fig:sketch}b).  The distance between the inner faces of the disks is $H=180 \ {\rm mm}$, which defines a flow volume of aspect ratio $H/R_c=1.8$. With the curved blades, the directions of rotation are no longer equivalent and we can either rotate the impellers anticlockwise ---with the convex face of the blades forward, direction~$(+)$ or clockwise (with the concave face of the blades forward, direction~$(-)$. 

The impellers are driven by two independent brushless $1.8 \ {\rm kW}$ motors, with speed servo loop control. The maximal torque they can reach is $11.5 \ {\rm N.m}$. The motor rotation frequencies $\{f_1 \ ; \ f_2\}$ can be varied independently in the range $1\leq f \leq 15 \ {\rm Hz}$. Below $1 \ {\rm Hz}$, the speed regulation is not efficient enough, and the dimensional quantities are measured with insufficient accuracy. We will consider for exact counter-rotating regimes $f_1=f_2$ the imposed speed of the impellers $f$. 

The experimental setup is thus axisymmetric and symmetric towards rotations of $\pi$ around any radial axis passing through the centre $O$ (\zob-symmetry), and we will consider here only \zob-symmetric mean solutions, though mean flows breaking this symmetry do exist for these impellers, at least at very high-Reynolds numbers \cite[]{ravelet2004}. A detailed study of the Reynolds number dependence of the \lq\lq global turbulent bifurcation\rq\rq ~is out of the scope of the present article and will be presented elsewhere. Also, since we drive the impellers independently, there is always a tiny difference between $f_1$ and $f_2$ and the \zob-symmetry of the system cannot be considered as exact. In the following, we will keep using this symmetry ---very useful to describe the observed patterns-- but we will keep in mind that our system is only an approximation of a \zob -symmetric system. The consequences on the dynamics will be analysed in the discussion (\S\,\ref{sec:disc:symmetries}).

\begin{table}
\begin{center}
\begin{tabular}{cccccc}
$C$ & $\mu$ at $15$\degre C & $\mu$ at $30$\degre C & $\rho$ & $Re$ range \\[3pt]
\hline
$99\%$ & $1700$ & $580$ & $1260$ & $50-2,000$\\
$93\%$ & $590$ & $210$ & $1240$ &  $130-5,600$\\
$85\%$ & $140$ & $60$ & $1220$ &  $550-19,000$\\
$81\%$ & $90$ & $41$ & $1210$ &  $840-28,000$\\
$74\%$ & $43$ & $20$ & $1190$ &  $1,800-56,000$\\
$0\%$ & $1.1$ & $0.8$ & $1000$ &  $570,000-1,200,000$\\
\end{tabular}
\caption{\label{tab:visco} Dynamic viscosity $\mu$ ($10^{-3} \ $Pa.s) at various temperatures, density $\rho$ (${\rm kg}.{\rm m}^{-3}$) at $20$\degre C and achievable Reynolds number range for various mass concentrations $C$ of glycerol in water.}
\end{center}
\end{table}

In the following, all lengths will be expressed in units of $R_c$. We also use cylindrical coordinates $\{r \ ; \ z\}$ and the origin is on the axis of the cylinder, and equidistant from the two impellers to take advantage of the \zob -symmetry (see figure~\ref{fig:sketch}a). The time unit is defined with the impeller rotation frequency $f$. The integral Reynolds number $Re$ is thus defined as $Re=2\pi f R_c^2 \nu^{-1}$ with $\nu$ the kinematic viscosity of the fluid. 

As in previous works \cite[]{marie04pof,ravelet2004,ravelet2005}, we use water at $20 - 30$\degre C as working fluid to get Reynolds numbers in the range $6.3 \times 10^4 \lesssim Re \lesssim 1.2 \times 10^6$. To decrease $Re$ down to laminar regimes, \ie to a few tens, we need a fluid with a kinematic viscosity a thousand times greater than that of water. We thus use $99\%$-pure glycerol which kinematic viscosity is $0.95 \times 10^{-3} \ {\rm m}^2.{\rm s}^{-1}$ at $20$\degre C \cite[]{handbook} and should be able to study the range $50 \lesssim Re\lesssim 900$. To cover a wide range of Reynolds numbers and match these two extreme ranges, we then use different mixes of glycerol and water, at temperatures between $15$ and $35$\degre C. The physical properties of these mixtures are given in table~\ref{tab:visco}, where $C$ is the mass percentage of glycerol in the mixture. Solutions samples are controlled in a Couette viscometer.

The temperature of the working fluid is measured with a platinum thermoresistance (Pt100) mounted on the cylinder wall ($\{r=1 \ ; \ z=0\}$). To control this temperature, thermoregulated water circulates in two heat exchangers placed behind the impellers. Plexiglas disks can be mounted between the impellers and heat exchangers to reduce the drag of the impellers-back-side flows. They are at typically $50$~mm from the impellers back sides. However, these disks reduce the thermal coupling: they are used in turbulent water flows and taken away at low Reynolds number. 

\subsection{Experimental tools, dimensionless measured quantities and experimental errors}
\label{subsec:mesure}

Several techniques have been used in parallel: flow visualisations with light sheets and air bubbles, torque measurements and velocity measurements.

Flow visualisations are made in vertical planes illuminated by approximately $2 \ {\rm mm}$ thick light sheets. We look at two different positions with respect to the flow: either the central meridian plane where the visualised components are the radial and axial ones or in a plane almost tangent to the cylinder wall where the azimuthal component dominates. Tiny air bubbles (less than 1 mm) are used as tracing particles. 

Torques are measured as an image of the current consumption in the motors given by the servo drives and have been calibrated by calorimetry. Brushless motors are known to generate electromagnetic noise, due to the Pulse-Width-Modulation supply. We use armoured cables and three-phase sinusoidal output filters (Schaffner FN5010-8-99), and the motors are enclosed in Faraday cages, which enhances the quality of the measurements. The minimal torques we measured are above $0.3 \ {\rm N.m}$, and we estimate the error in the measurements to be $\pm 0.1 \ {\rm N.m}$. The torques $T$ will be presented in the dimensionless form:
$$K_p=T \ (\rho R_c^5 (2\pi f)^2)^{-1}$$

Velocity fields are measured by Laser Doppler Velocimetry (LDV). We use a single component DANTEC device, with a He--Ne Flowlite Laser (wave length $632.8 \ {\rm nm}$) and a BSA57N20 Enhanced Burst Spectrum Analyser. The geometry of the experiment allows us to measure in one point either the axial component $V_z(t)$ or the azimuthal component $V_{\theta}(t)$. Though the time-averaged velocity field $\bf V$ is not a solution of the Navier-Stokes equations, it is a solenoidal vector field, and it is axisymmetric. We thus use the incompressibility condition $\nabla \cdot {\bf V} = 0$ to compute the remaining radial component $V_r$. 

The measurements of the time-averaged velocity field are performed on a $\{r \times z\}=11 \times 17$ grid, using a weighting of velocities by the particles transit time, to get rid of velocity biases as explained by \citet{buchhave79}. This acquisition mode does not have a constant acquisition rate, so we use a different method for the acquisition of well-sampled signals to perform temporal analysis at single points. In this so-called dead-time mode, we ensure an average data rate of approximately $5 \ {\rm kHz}$, and the Burst Spectrum Analyser takes one sample every single millisecond such that the final data rate is $1 \ {\rm kHz}$. For practical reasons, this method is well-suited for points close to the cylindrical wall, so we choose the point $\{r=0.9 \ ; \ z=0\}$ for the measurements in figures~\ref{fig:ldvprecision1}, \ref{fig:ldvprecision2} and \ref{fig:niveaubruitdere}. The signals are re-sampled at $300$Hz by a \lq\lq Sample And Hold\rq\rq~ algorithm \cite[]{buchhave79}.

Let us now consider the experimental error on the Reynolds number value. The speed servo-loop control ensures a precision of $0.5\%$ on $f$, and an absolute precision of $\pm0.002$ on the relative difference of the impellers speeds $(f_1-f_2)/(f_1+f_2)$. The main error on the Reynolds number is thus a systematic error that comes from the estimation of the viscosity. As far as the variation of the viscosity with temperature is about $4\%$ for $1$\degre C and the variation with concentration is about $5\%$ for $1\%$ of mass concentration, we estimate the absolute error on $Re$ to $\pm 10\%$ (the temperature is known within $1$\degre C). However, the experimental reproducibility of the Reynolds number is much higher than $\pm 10\%$. In the range $100 \lesssim Re \lesssim 500$ we are able to impose $Re$ within $\pm 5$.

\section{From order to turbulence: description of the regimes}
\label{sec:descr_regimes}

This section describes the evolution of the flow from the laminar regime to the fully developed turbulence, \ie for $30 \lesssim Re \lesssim 1.2 \times 10^{6}$. This wide-range study has been carried for the negative sense of rotation $(-)$ of the propellers. 

\subsection{Basic state at very low Reynolds number} 
\label{sec:basic}

\begin{figure*}
\begin{center}
(a) \includegraphics[clip,height=38mm]{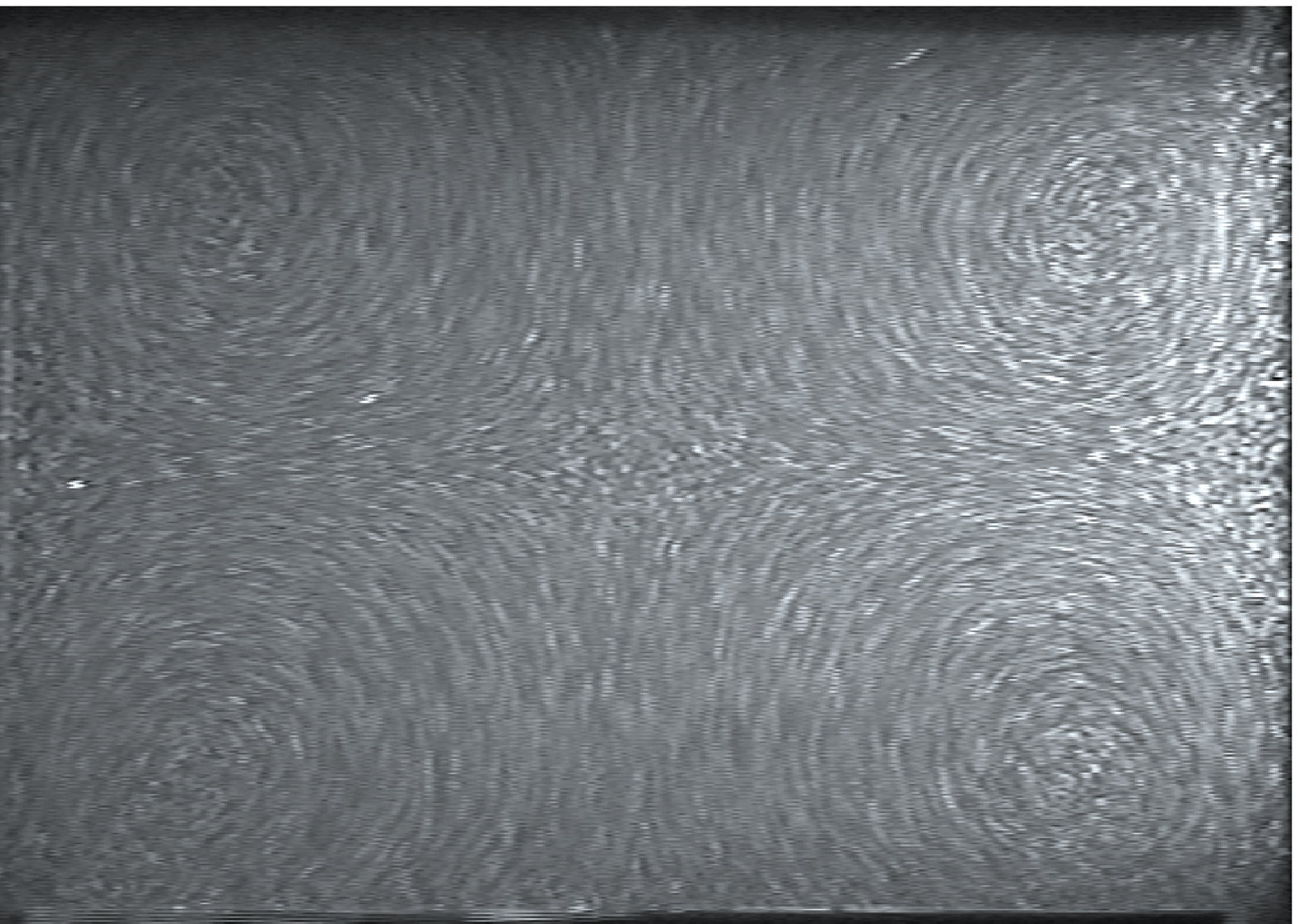}
\hspace{0.5cm}
\includegraphics[clip,height=38mm]{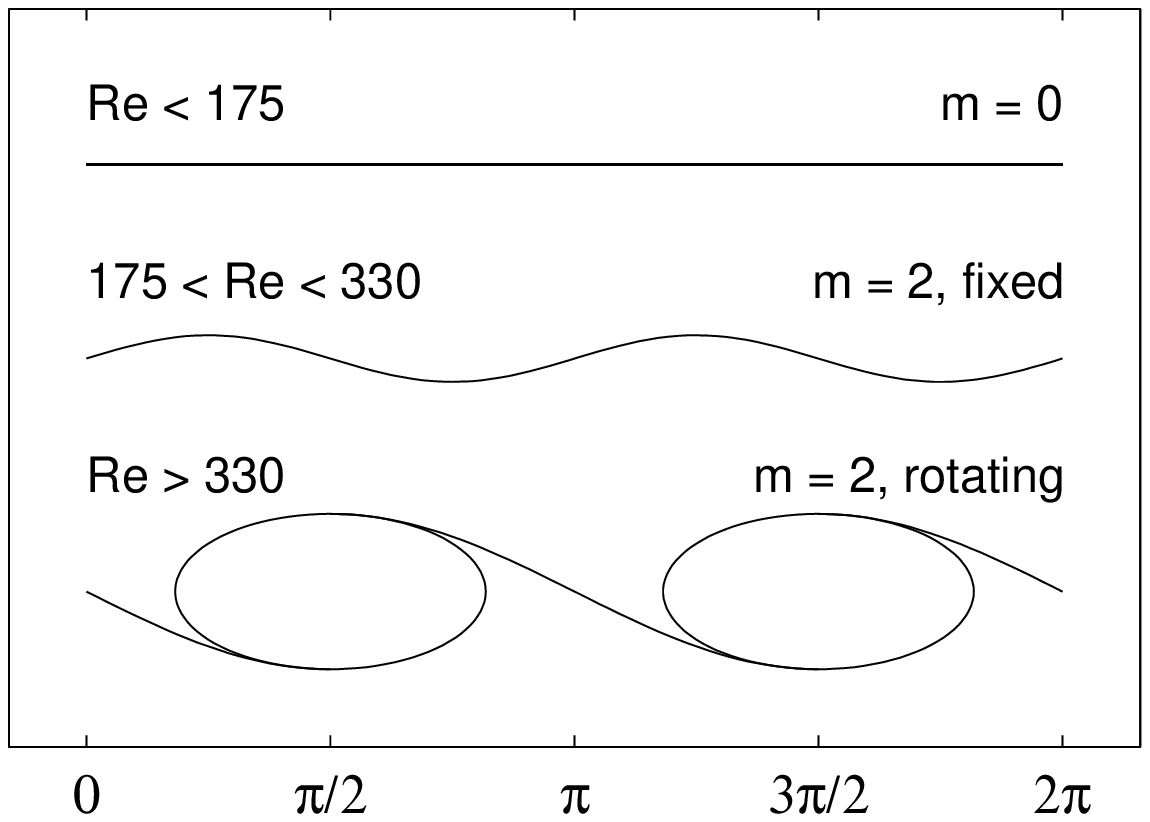} (e)\\
\vspace{0.3cm}
\includegraphics[clip,height=38mm]{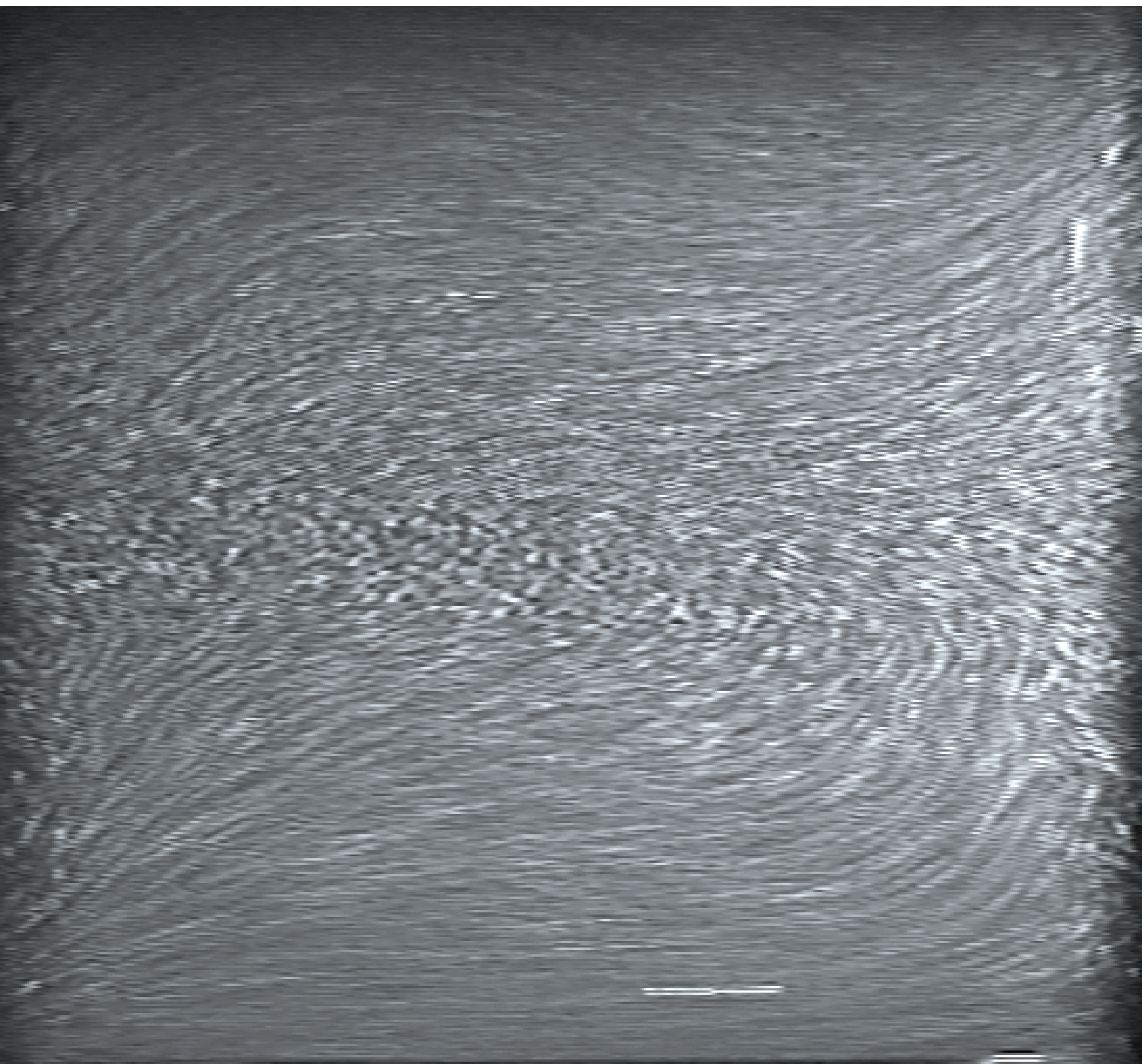}
\hspace{0.05cm}
\includegraphics[clip,height=38mm]{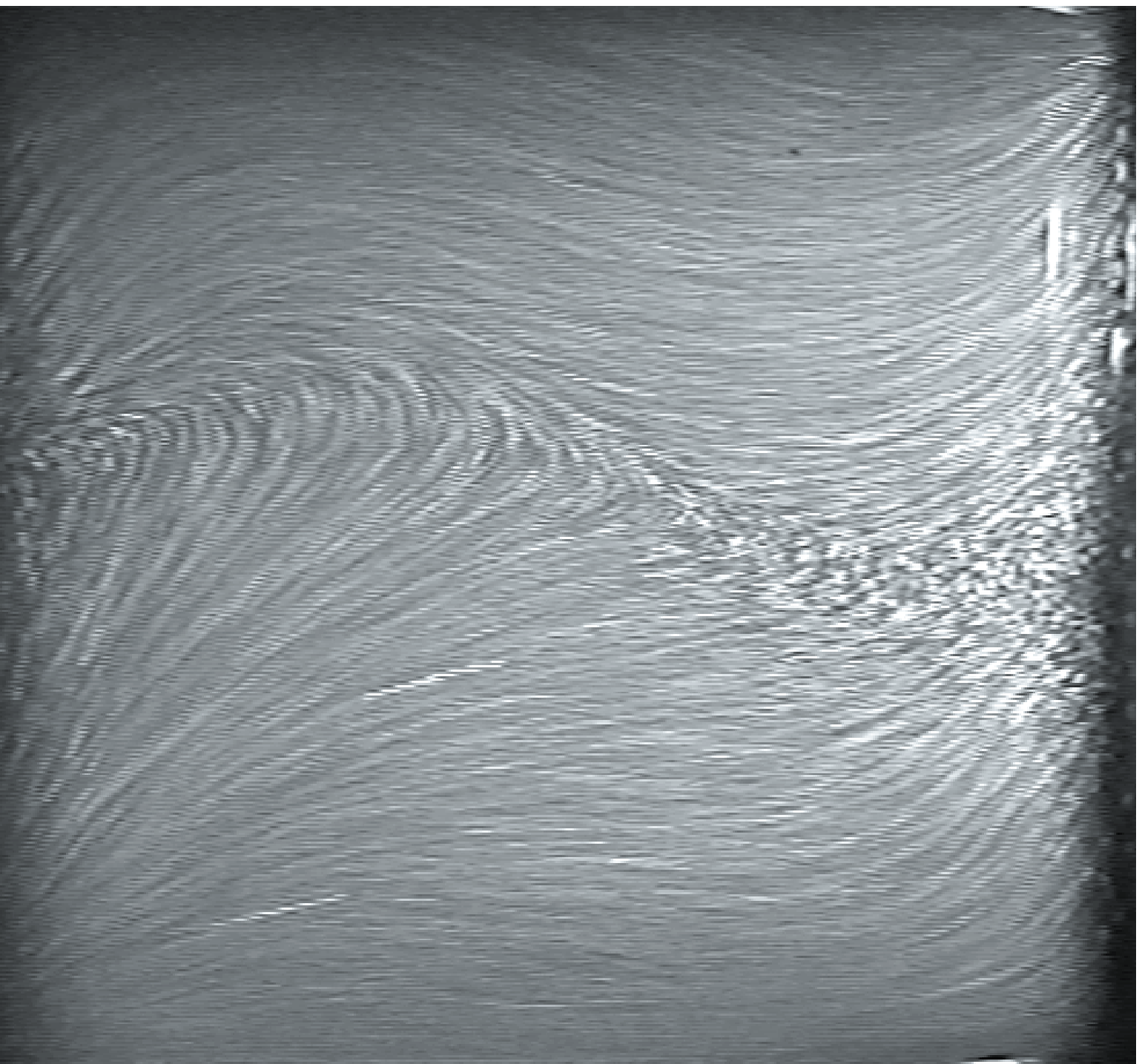}
\hspace{0.05cm}
\includegraphics[clip,height=38mm]{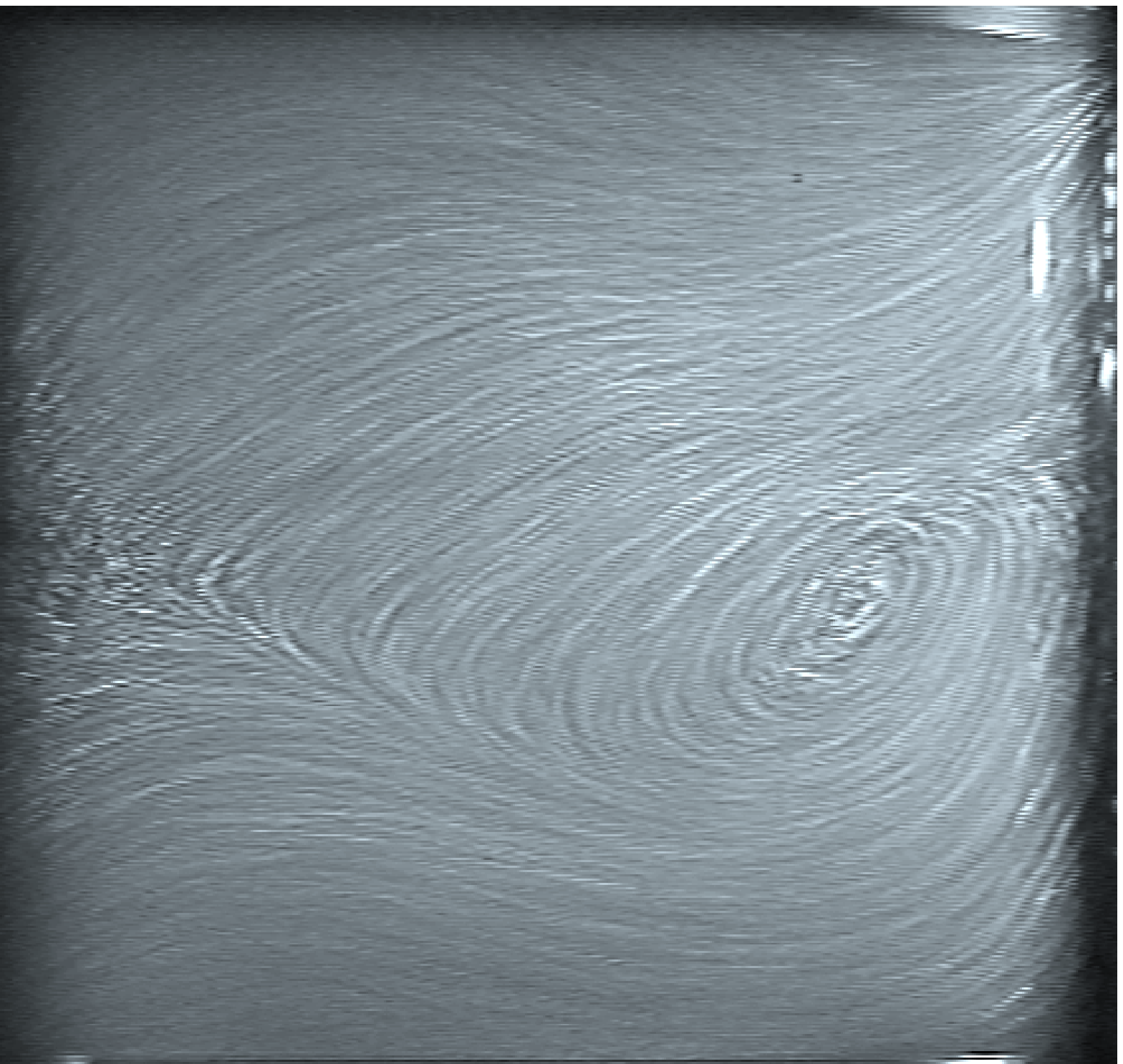}\\
(b) \hfill (c) \hfill (d)
\caption{Visualisation and schematics of the basic laminar flow for impellers rotating in direction $(-)$. The lightning is made with a vertical light sheet. Pictures are integrated over $1/25 \ {\rm s}$ with a video camera, and small air bubbles are used as tracers. Picture height is $H-2h = 1.4 R_c$. Laminar axisymmetric flow at $Re=90$, meridian view (a). Views in a plane near the cylinder wall at $Re=90$ (b), $Re=185$ (c) and $Re=345$ (d). The development of the first $m=2$ instabilities ---steady undulation (c) and rotating vortices (d)--- is clearly visible on the shape of the shear-layer. We give sketches of the shear-layer for these Reynolds numbers in (e).}
\label{fig:visus}
\end{center}
\end{figure*}

At very low-Reynolds number, the basic laminar flow respects the symmetries of the problem. It is stationary, axisymmetric and \zob -symmetric. This state is stable at $Re=90$, where we present a flow visualisation in figure~\ref{fig:visus}(a-b). In figure~\ref{fig:visus}(a), the light sheet passes through the axis of the cylinder. The visualised velocities are the radial and axial components. The poloidal part of the flow consists of two toric recirculation cells, with axial pumping directed to the impellers. 

The flow is also made of two counter-rotating cells, separated by an azimuthal flat shear-layer, which can be seen in figure~\ref{fig:visus}(b) where the light sheet is quasi-tangent to the cylinder wall. Both the azimuthal and axial component vanishes in the plane $z=0$ which is consistent with the axisymmetry and the \zob -symmetry. This flat shear-layer is sketched in figure~\ref{fig:visus}(e). A LDV velocity field is presented in section~\ref{sec:visc2inert} (figure~\ref{fig:comparaisoncaro}c-d).

\subsection{First instability -- stationary bifurcation} 
\label{sec:m=2}

The first instability for this flow has been determined by visualisation and occurs at $Re = 175 \pm 5$ for both directions of rotation. The bifurcation is supercritical, non-hysteretic, and leads to a stationary regime, with an azimuthal modulation of $m=2$ wave number. We present a visualisation of this secondary state in figure~\ref{fig:visus}(c), at $Re=185$. The axisymmetry is broken: one can see the $m=2$ modulation of the shear-layer, also sketched in figure~\ref{fig:visus}(e). One can also note that \zob-symmetry is partly broken: the bifurcated flow is \zob -symmetric with respect to two orthogonal radial axis only. This first instability is very similar to the Kelvin--Helmholtz instability. \cite{nore03} made a proper theoretical extension of the Kelvin--Helmholtz instability in a cylinder. Their model is based on the use of local shear-layer thicknesses and Reynolds numbers to take into account the radial variations in the cylindrical case.

We observe this $m=2$ shear-layer to rotate very slowly in a given direction with a period $7500 f^{-1}$. This corresponds to a very low frequency, always smaller than the maximum measured dissymmetry of the speed servo loop control between both independent motors (\S\,\ref{subsec:mesure}). This is probably the limit of the symmetry of our system, \ie the pattern is at rest in the slowly rotating frame where both frequencies are strictly equal (see discussion in~\S\,\ref{sec:disc:symmetries}). For convenience, we will describe the dynamics in this frame.

The laminar $m=2$ stationary shear-layer pattern is observed up to typically $Re\simeq 300$ where time-dependence arises.

\label{sec:states}

To investigate the time-dependent regimes, we now also perform precise velocity measurements at a given point in the shear-layer. We measure the azimuthal component $v_{\theta}$ at $\{r=0.9 \ ; \ z=0\}$, using the dead-time acquisition mode (see~\S\,\ref{sec:exp}). 

Below, we describe and illustrate the observed dynamics and the building-up of the chaotic and turbulent spectra. The next section \ref{sec:quant} is complementary: we quantitatively characterize the transitions as much as we can, we discuss the mechanisms and we finally propose a global supercritical view of the transition to turbulence.

\subsection{From drifting patterns to chaos}
\label{sec:pres:low_freq_spectr}

\begin{figure*}
\begin{center}
\includegraphics[clip,width=130mm]{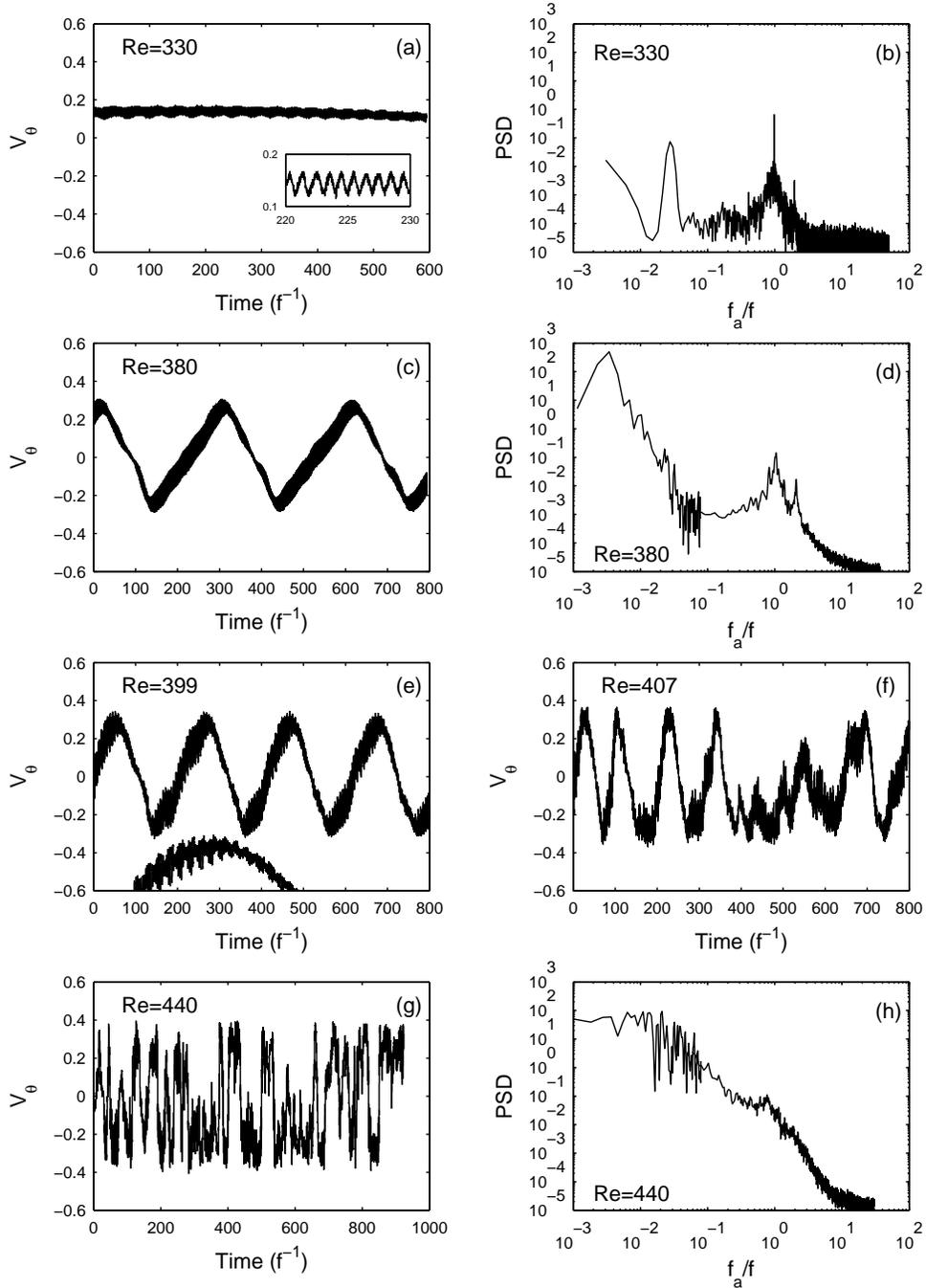}
\caption{Temporal signals $v_{\theta}(t)$ measured by LDV at $\{r=0.9 \ ; \ z=0\}$ and power spectral densities (PSD), at: (a-b)~$Re=330$, (c-d)~$Re=380$, (e)~$Re=399$~, (f)~$Re=408$ and (g-h)~$Re=440$. $f_a$ is the analysis frequency whereas $f$ is the impellers rotation frequency. Inset in (a): zoom over the fast oscillation at frequency $f$.  In (e), a small part of the signal is presented with time magnification ($\times 4$) and arbitrary shift to highlight the modulation at $6.2 f^{-1}$. Power spectra are computed by the Welch periodogram method twice: with a very long window to catch the slow temporal dynamics and with a shorter window to reduce fast scales noise.}
\label{fig:ldvprecision1}
\end{center}
\end{figure*}

We present time series of the velocity and power spectral densities at five Reynolds numbers in figure~\ref{fig:ldvprecision1}: $Re=330 \pm 5$, $ Re=380 \pm 5$, $Re=399 \pm 5$, $ Re=408 \pm 5$ and $ Re=440 \pm 5$.

\subsubsection{Oscillation at impeller frequency}
\label{sec:pres:f}

The point at $Re=330 \pm 5$ is the first point where a clear temporal dynamics is observed: a sharp peak in the spectrum (figure~\ref{fig:ldvprecision1}b) is present at the impeller rotation frequency $f_a=f$ ---emphasized in the Inset of figure~\ref{fig:ldvprecision1}(a). This oscillation exists for higher $Re$ with the same small amplitude: it is too fast to be explicitly visible on the long time series of figure~\ref{fig:ldvprecision1}, but it is responsible for the large width of the signal line.

In comparison, a similar measurement performed at $Re \simeq 260$ reveals a flat signal with a very low flat-spectrum with just a tiny peak ---$1/1000$ of the amplitude measured at $Re=330$--- at $f_a=f$ and we have no data in-between to check an evolution.

On the spectra, we observe the first harmonic, but never shows the expected blade frequency $8f$ nor a multiple. So, it is not clear if it corresponds to the basic fluid instability mode or just to a small precessing mode due to the misaligning of the impeller axis or even to mechanical vibrations transmitted to the fluid through the bearings. Since the travelling-wave mode of the next paragraph is much stronger and richer in dynamics we will consider that the signal at $f_a=f$ is a \lq\lq minor\rq\rq ~phenomenon, \ie a perturbation of the steady $m=2$ mode.

In figure~\ref{fig:ldvprecision1}(a) the mean velocity is not zero, but around $v_{\theta}=+0.17$ during the $600$ time units of acquisition, \ie during $600$ disks rotations. This value of the velocity has no special meaning and depends on the phase between the fixed measurement point and the slowly drifting shear-layer (\S~\ref{sec:m=2}). The measurement point indeed stays on the same side of the shear-layer for this time-series but, on much longer time scales, we measure the $m=2$ shear-layer rotation typical period. 

Further observation on signal and spectrum of figure~\ref{fig:ldvprecision1}a-b reveals some energy at low frequency around $f_a \simeq f/30$, corresponding to slowly relaxing modulations: the slowness of this relaxation is the clear signature of the proximity of a critical point.

\subsubsection{Drifting/Travelling Waves}
\label{sec:pres:TW}

For $330 < Re < 389$ the velocity signal is periodic with a low frequency $f_D$. This is illustrated at $Re=380$ in figure~\ref{fig:ldvprecision1}(c-d). The mean velocity is now zero: the shear-layer rotates slowly such that the measurement point is alternatively in the cell rotating with the upper impeller ($v_{\theta}>0$) and in the cell rotating with the lower impeller ($v_{\theta}<0$). Visualisations confirm that this corresponds to a travelling wave (TW) or a drifting pattern and also show that the $m=2$ shear-layer is now composed of two vortices (figure~\ref{fig:visus}d) and thus deserves the name \lq\lq mixing-layer\rq\rq. Along the equatorial line, one notice that the parity is broken or the vortices are tilted \cite[]{coullet90,knobloch96}. The velocity varies between $-0.3 \lesssim v_{\theta} \lesssim 0.3$. The drift is still slow but one order of magnitude faster than the drift described above for the \lq\lq steady\rq\rq ~$m=2$ pattern: one can see two periods during $600$ time units, \ie $f_D=f/300$ which is very difficult to resolve by spectral analysis owing to the shortness of the signal ({\em cf.} caption of figure~\ref{fig:ldvprecision1}). At $Re=380$ (figure~\ref{fig:ldvprecision1}c-d), the peak at the rotation frequency is still present, but starts to spread and becomes broadband. The power spectral density at frequencies higher than $3f$ decreases extremely rapidly to the noise level. Let us note that \zob-symmetry remains only with respect to a pair of orthogonal radial axis which rotates with the propagating wave.

\subsubsection{Modulated Travelling Waves}
\label{sec:pres:MTW}

For $389 < Re < 408$ the signal reveals quasiperiodicity, \ie modulated travelling waves (MTW), shown in figure~\ref{fig:ldvprecision1}e-f at $Re=399$ and $408$. The MTW are regular, \ie strictly quasiperiodic below $Re=400$ and irregular above. The modulation frequency ---see magnified ($\times 4$) piece of signal in figure~\ref{fig:ldvprecision1}e--- is $f_M = f/(6.2\pm0.2)$ whatever $Re$, even above $Re=400$. It is much faster than the drift frequency ($f_D \sim f/200$) and seems to be related to oscillations of the mixing-layer vortex cores on the movies.

\subsubsection{Chaotic regime}
\label{sec:pres:chaos}

The upper limit of the regular dynamics is precisely and reproducibly $Re=400$ and there is no hysteresis. From the visualisations, we observe that the $m=2$ symmetry is now broken. The mixing-layer vortices, which are still globally rotating around the cell in the previous direction, also behave more and more erratically with increasing $Re$: their individual dynamics includes excursions in the opposite direction as well as towards one or the other impeller. The velocity signal also looses its regularity (see figure~\ref{fig:ldvprecision1}e-g).

When this disordered regime is well established, \emph{e.g.} for $Re=440$  (figure~\ref{fig:ldvprecision1}g), it can be described as series of almost random and fast jumps from one side to the other side of the $v=0$ axis. The peaks reached by the velocity are now in the range $-0.4 \lesssim v_{\theta} \lesssim 0.4$. The spectral analysis of the signal at $Re=440$ (figure~\ref{fig:ldvprecision1}h), does not reveal any well-defined frequency peak any more. However, a continuum of highly-energetic fluctuations at low frequency, below $f_a=f$ and down to $f_a=f/100$, emerges. A small bump at the rotation frequency $f$ is still visible, and a region of fast fluctuations above the injection frequency also seems to arise. Although we did not carried detailed Poincar\'e analysis or equivalent and cannot characterize clearly a scenario, we find this transition and this regime typical enough to call it \lq\lq chaos\rq\rq ~(see also \ref{sec:quant:order2chaos}).

\begin{figure*}
\begin{center}
\includegraphics[clip,width=130mm]{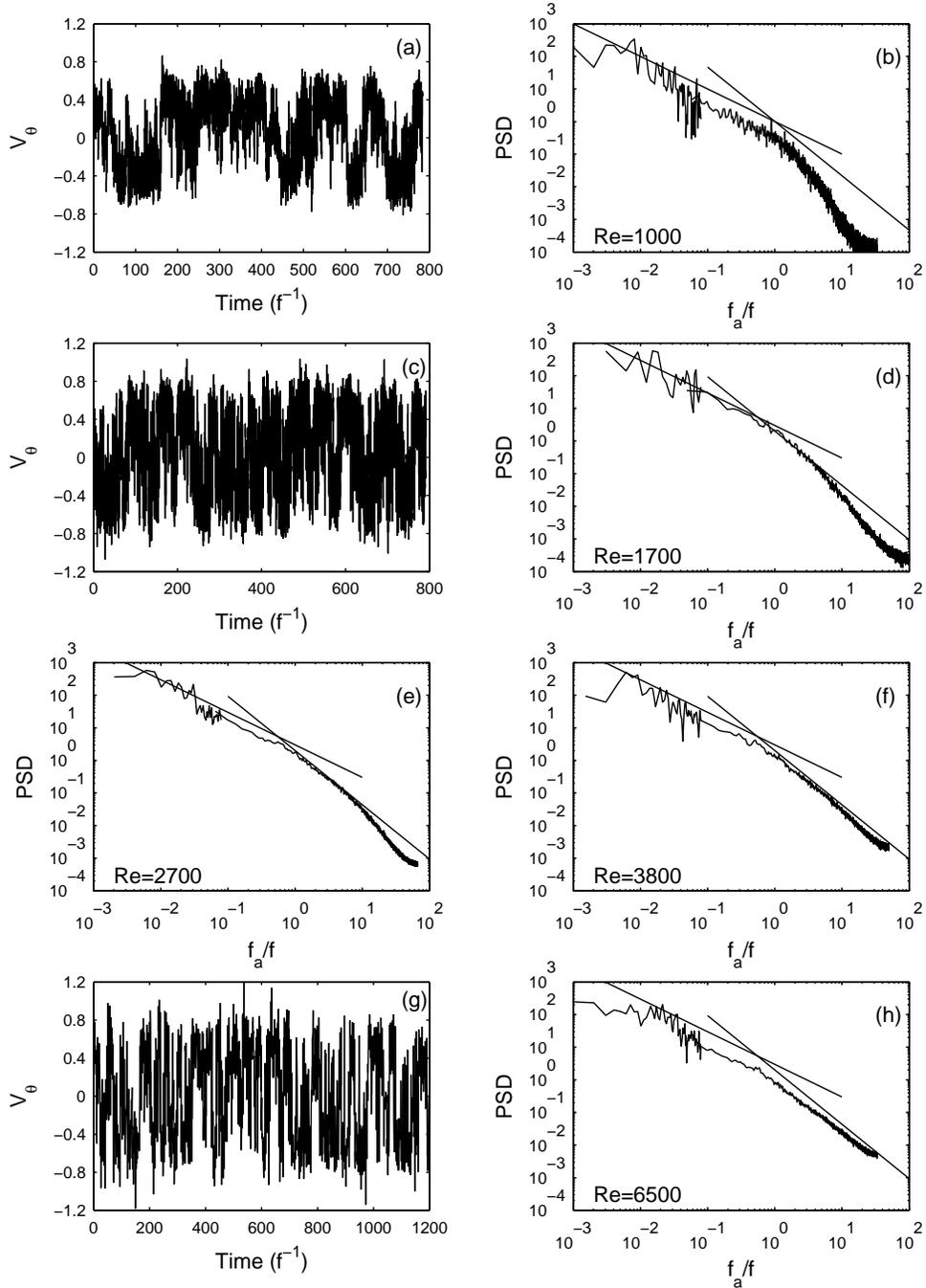}
\caption{(a-b): Temporal signal $v_{\theta}(t)$ measured by LDV at $\{r=0.9 \ ; \ z=0\}$ and power spectral density at $Re=1.0 \times 10^3$. (c-d): Temporal signal and power spectral density at $Re=1.7\times 10^3$. (e-f): Power spectral densities at $2.7\times 10^3$ and $3.8\times 10^3$. (g-h): Temporal signal and power spectral density at $6.5\times 10^3$. Solid lines in the power spectra plots are power-law eye-guides of slope $-1$ and $-5/3$. Spectra are computed as explained in the caption of figure~\ref{fig:ldvprecision1}.}
\label{fig:ldvprecision2}
\end{center}
\end{figure*}

\subsection{From chaos to turbulence: building a continuous spectrum}
\label{sec:pres:high_freq_spectr}

Increasing further the Reynolds number, one obtains the situation depicted in figure~\ref{fig:ldvprecision2}. The time-spectrum is now continuous but still evolving. We describe the two parts, below and above the impellers frequency $f$.

\subsubsection{Slow time scales}
The slow dynamics which has already been described at $Re=440$ (figure~\ref{fig:ldvprecision1}g,h) could be thought as depending only on the largest spatial scales of the flow. It is well built above $Re \simeq 10^3$ (figure~\ref{fig:ldvprecision1}b). The mean velocities corresponding to each side of the mixing layer are of the order of $\pm0.6$ at $Re=1.0 \times 10^3$ and above (figures~\ref{fig:ldvprecision2}a, c and g). The power spectral density below the injection frequency seems to behave with a $f^{-1}$ power-law over two decades (see discussion \S~\ref{sec:disc:3scales}) for all these Reynolds numbers (figure~\ref{fig:ldvprecision2}). The spectral density saturates below $10^{-2}f$.

\subsubsection{Fast time scales}
However, the fast time scales, usually interpreted as a trace of small spatial scales fluctuations, evolve between $Re=1.0 \times 10^3$ and $Re=6.5\times 10^3$. At the former Reynolds number, there are few fast fluctuations decaying much faster than $f^{-5/3}$ (figure~\ref{fig:ldvprecision2}(b)) and the intermittent changeovers are easy to identify in the temporal signal in figure~\ref{fig:ldvprecision2}(a). If Reynolds number is increased, the fast (small scales) fluctuations get bigger and bigger and behave as a power law over a growing frequency range (figure~\ref{fig:ldvprecision2}e, f, h). The measured slope is of order of $-1.55$ over $1.5$ decade, \ie $10\%$ lower than the classical $f^{-5/3}$. This value of the exponent could be ascribed to the so-called \lq\lq bottleneck\rq\rq ~effect \cite[]{falkovich1994} and is compatible with the values given by \citet{lohse1995} ($-1.56\pm 0.01$) for a Taylor-micro-scale Reynolds number $R_{\lambda} \simeq 100$, which is an estimation for our flow based on the results of \citet{zocchi94}.

\section{Quantitative characterization of the transitions}
\label{sec:quant}

The various dynamic states encountered have been described and illustrated in the previous section. Now, we wish to analyse some characteristic measurements ---the amplitude of the velocity fluctuations and their main frequencies---, extract thresholds and critical behaviours and then address the question of the nature of the reported transitions.

\subsection{From order to turbulence: a global supercriticality}
\label{sec:quant:supercrit}

\begin{figure}
\begin{center}
\includegraphics[clip,width=80mm]{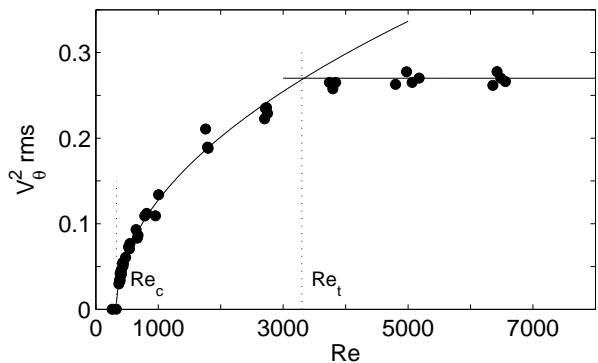}
\caption{Variance of $v_{\theta}(t)$ measured at $\{r=0.9 \ ; \ z=0 \}$ \vs $Re$. Solid line: non-linear fit of the form $v_{\theta \ rms}^2=a \times (Re-Re_c)^{1/2}$, fitted between $Re=350$ and $Re=2500$. The regression coefficient is $R^2=0.990$, and the fit gives $Re_c=328 \pm 8$ with $95\%$ confidence interval. The intersection between this fit and the asymptotic value $v_{\theta \ rms}^2 \simeq 0.27$ gives $Re_t=3.3 \times 10^3$.}
\label{fig:niveaubruitdere}
\end{center}
\end{figure}

It is known that fully turbulent von K\'arm\'an flow can generate velocity fluctuations of typically $50 \%$ of the driving impellers velocity. So, we compute the variance $v_{\theta \ rms}^2$ of the LDV-time-series {\it versus} the Reynolds number. This quantity is homogeneous to a kinetic energy and may be referred to as the azimuthal kinetic energy fluctuations in the mixing layer. With this method, we consider altogether the broadband frequency response of the signal. The results are reported in figure~\ref{fig:niveaubruitdere} for all the measurements performed between $260 \lesssim Re \lesssim 6500$. 
Except at time-dependence threshold, this quantity behaves very smoothly: it can be fitted between $Re=350$ and $Re=2500$ with a law in the square root of the distance to a threshold $Re_c\simeq330$ (figure~\ref{fig:niveaubruitdere}):
$$v_{\theta \ rms}^2 \propto (Re-Re_c)^{1/2}.$$

Since we will show below that $Re_c$ is precisely the threshold for time-dependence, we can make here the hypothesis that $v_{\theta \ rms}^2$ is a global order parameter for the transition to turbulence, \ie for the transition from steady flow to turbulent flow taken as a whole. With this point of view the transition is globally supercritical.

\subsection{Transitions from order to chaos}
\label{sec:quant:order2chaos}

\begin{figure}
\begin{center}
\includegraphics[clip,width=80mm]{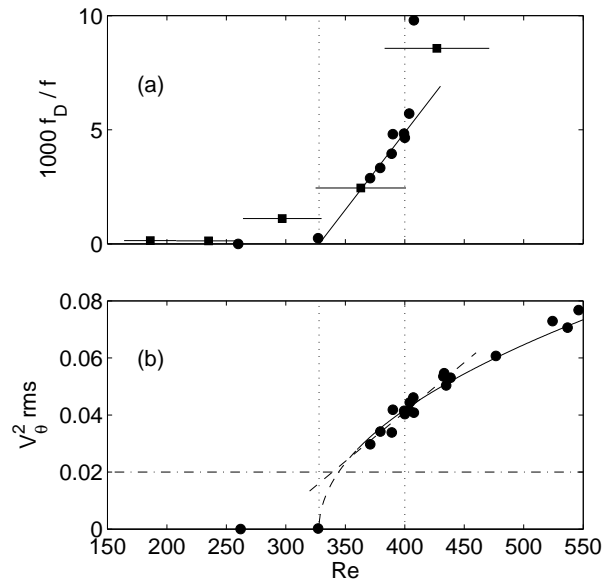}
\caption{(a) Low-frequency $f_{D}$ of the quasi-periodic regime of velocity $v_{\theta}(t)$ measured at $\{r=0.9 \ ; \ z=0\}$ (circles) and drift frequency of the $m=2$ shear-layer pattern from flow visualisations (squares with high horizontal error bars due to poorer temperature control). The solid line is a linear fit of $f_{D}$ between the two thresholds $Re_{TW}=330$ and $Re_{chaos}=400$, indicated by vertical dotted lines. (b) Zoom of figure~\ref{fig:niveaubruitdere}. The dashed line is a linear fit of the lowest data between $Re=350$ and $Re=450$. Close to the threshold, it crosses the dash-dotted line which corresponds to the velocity due to the drift and estimates the level of imperfection.
}

\label{fig:QP}
\end{center}
\end{figure}

We now turn to the very first steps of the transition to time dependence. We monitor the main frequencies of the mixing layer dynamics in the TW- and MTW-regimes (see \S\,\ref{sec:pres:low_freq_spectr}). In these regimes, even if only few periods are monitored along single time-series, we carefully estimate the period by measuring the time delay between crossings of the $v=0$ axis. These value are reported in figure~\ref{fig:QP}(a) with circles. In an equivalent way, the periodicity of the travelling of the mixing-layer vortices on the visualisations give complementary data, represented by squares on the same figure. 

\subsubsection{Onset of time-dependence}
\label{sec:quant:330}

The drift frequency $f_{D}$ of the travelling waves behaves linearly with $Re$ above a threshold $Re_{TW}$ very close to $330$. Both measurement methods agree even if the visualisations deserves large error bars in $Re$ at least due to the shortness of our records and to a poorer thermal control. So, the fit is made on velocity data only. We observe some level of imperfection in the quasi-periodic bifurcation, due to the pre-existing slow drift below $Re_{TW}$: we always observe the mixing-layer to start rotating in the sense of the initial drift. 

We show in figure~\ref{fig:QP}(b) a zoom of figure~\ref{fig:niveaubruitdere}, \ie the amplitude of the kinetic energy fluctuations. We observe that both quadratic amplitude fit and linear frequency fit converge to exactly the same threshold $Re_{TW}=Re_c=328$. We can conclude that the low-frequency mode at $f_a=f_{D}$ bifurcates at $Re=330\pm5$ through a zero-frequency bifurcation for $f_{D}$.

The question is thus how the amplitude precisely behaves at onset. There is obviously a lack of data in the narrow range $300 \lesssim Re \lesssim350$ (figure~\ref{fig:QP}(b)). It is due to the high temperature dependence of the viscosity in this regime (Reynolds varied quite fast even with thermal control) and to some data loss at the time of experimental runs.  Despite this lack, we present these observations because of the consistency of the different types of data ---visualisations, LDV, torques--- over the wide Reynolds number range. The horizontal line $v_{\theta \ rms}^2=0.02$ in figure~\ref{fig:QP}(b) corresponds to an amplitude of typically $0.15$ for $v_{\theta}$, which is produced just by the initial shear-layer drift (see the maximum speed in figure~\ref{fig:ldvprecision1}a). This value is in good agreement with a linear extrapolation over the lower range of figure~\ref{fig:QP}(b) and thus again with an imperfect bifurcation due to the drift. If we reduce the drift by better motor frequencies matching, the onset value of $v_{\theta}$ will depend on the longitude of the probe location and the parabola of figure~\ref{fig:QP}(b) could perhaps be observed on the $m=2$ shear-layer nodes.

\subsubsection{Transition to chaos}
\label{sec:quant:400}

The transition to chaos is very sharply observed for $Re > Re_{chaos}=400$. There is no hysteresis. Just above the chaotic threshold in the MTW regime (figure~\ref{fig:ldvprecision1}f), the signal sometimes exhibits a few almost-quasi-periodic oscillations still allowing us to measure a characteristic frequency. The measured values have been also plotted on figure~\ref{fig:QP}(a) and are clearly above the linear fit. This could reveal a vanishing time scale, \ie a precursor for the very sharp positive/negative jumps of $v_{\theta}$ reported in the chaotic and turbulent regimes. 

We do not clearly observe any evidence of mode locking between the present frequencies which are in the progression $f/200 \rightarrow f/6.2 \rightarrow f$ and there is no trace of sub-harmonic cascade on any of each. This could be linked to a three-frequency scenario \emph{\`a la} Ruelle--Takens \cite[]{manneville1990}.

\subsection{Transition to full turbulence}
\label{sec:fullturb}

\subsubsection{Torque data}
\label{sec:quant:torque}

\begin{figure}
\begin{center}
\includegraphics[clip,width=85mm]{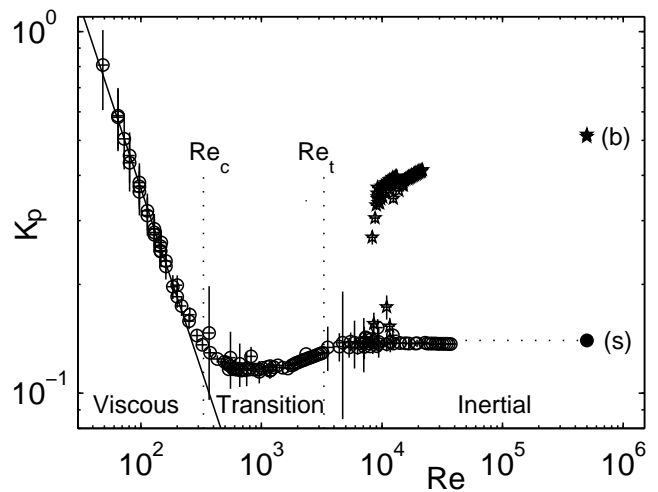}
\caption{Dimensionless torque $K_p$ \vs $Re$ in a log-log scale for the negative sense of rotation $(-)$ of the impellers.
The main data ($\circ$) corresponds to the symmetric (s)-flow regime described in this part of the article. For completeness, the high-torque branch ($\star$) for $Re \gtrsim 10^4$ corresponds to the (b)-flow regime \cite[]{ravelet2004}, \ie to the \lq\lq turbulent bifurcation\rq\rq ~(see \S~\ref{sec:bifglob}). Since both motors do not deliver the same torque in this \zob-symmetry broken (b)-flow, the average of both values is plotted.
Relative error on $Re$ is $\pm10\%$~; absolute error of $\pm0.1 \ {\rm N.m}$ on the torque. $Re_c$ and $Re_t$ are the transition values computed from the fits of figure~\ref{fig:niveaubruitdere}. 
The single points, displayed at $Re=5 \times 10^5$, correspond to measurements in water, where $K_p$ is extracted from a fit of the dimensional torque in $a+b\times f^2$ for $2 \times 10^5 \lesssim Re \lesssim 9 \times 10^5$ \cite[]{ravelet2005}.}
\label{fig:kpdere_anti}
\end{center}
\end{figure}

Complementary to the local velocity data, information can be collected on spatially integrated energetic data, \ie on torque measurements $K_p(Re)$ (figure~\ref{fig:kpdere_anti}). The low-Reynolds viscous part will be described below (\S~\ref{sec:visc2inert}) as well as the high-torque bifurcated branch (\S~\ref{sec:bifglob}). In the high-Reynolds number regimes, the torque reaches an absolute minimum for $Re \simeq 1000$ and becomes independent of $Re$ above $3300$.

\subsubsection{From chaos to turbulence}
\label{sec:quant:1000}

Is there a way to quantitatively characterize the transition or the crossover between chaos and turbulence? It seems to be no evidence of any special sign to discriminate between the two regimes. An empirical criterion we could propose would be the completeness of the $(f_a/f)^{-1}$ low-frequency part of the spectrum, clearly achieved for $Re=1000$ (figure~\ref{fig:ldvprecision2}b). This region also corresponds to the minimum of the $K_p(Re)$ curve (figure~\ref{fig:kpdere_anti}). One can propose that below this Reynolds number, the power injected at the impeller rotation frequency mainly excites low frequencies belonging to the \lq\lq chaotic\rq\rq~ spectrum, whereas above $Re \simeq 1000$ it also drives the high frequencies through the Kolmogorov-Richardson energy cascade.

\subsubsection{Inertial turbulence}
\label{sec:quant:3300}

\begin{figure}
\begin{center}
\includegraphics[clip,width=80mm]{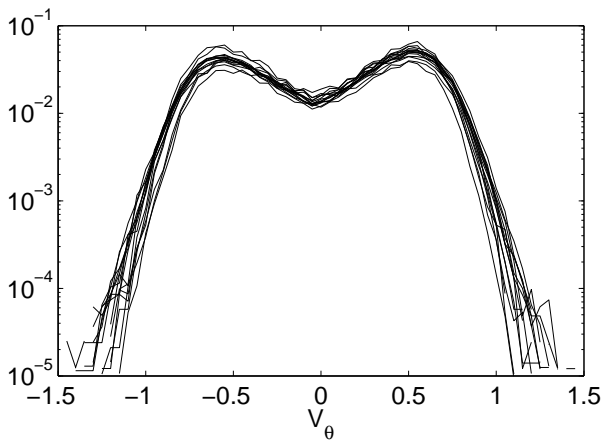}
\caption{Probability density function (PDF) of $v_{\theta}$ for $16$ Reynolds numbers in the range $2.5 \times 10^3 \lesssim Re \lesssim 6.5 \times 10^3$.}
\label{fig:PDFchameau}
\end{center}
\end{figure}

The $(Re-Re_c)^{1/2}$ behaviour can be fitted through the quasi-periodic and chaotic regimes, up to $Re \sim 3000$. Here, the azimuthal kinetic energy fluctuations level saturates at $v_{\theta \ rms}^2 \simeq 0.27$, \ie fluctuations of velocities at this point of the mixing-layer are of the order of $50 \%$ of the impeller tip speed. This saturation is also revealed by the Probability Density Functions (PDF) of $v_{\theta}$ presented in figure~\ref{fig:PDFchameau}. These PDF are computed for $16$ Reynolds numbers in the range $2.5 \times 10^3 \lesssim Re \lesssim 6.5 \times 10^3$. One can notice the bimodal character of the PDF: the two bumps, which are symmetric, correspond to the two counter-rotating cells. Furthermore, all these PDF collapse and are therefore almost independent of $Re$ in this range. This is also consistent with the spectral data of figure~\ref{fig:ldvprecision2}(b-d) where the $(f_a/f)^{-1}$ slowest time-scales regions which contain most of the energy ---below $f$--- appear similar for $Re=1.0\times 10^3$ and above (figure~\ref{fig:ldvprecision2}). The crossover Reynolds number $Re_t$ at which the kinetic energy of fluctuations saturates in figure~\ref{fig:niveaubruitdere} is estimated by taking the intersection of the horizontal asymptote with the fit: $Re_t=3.3 \times 10^3$. This value corresponds precisely to the value where the asymptotic plateau is reached in the $K_p$ \vs $Re$ diagram (figure~\ref{fig:kpdere_anti}). In such an inertially-driven turbulent flow, the bulk dissipation is much stronger than the dissipation in boundary layers and the global dimensionless quantities thus do not depend on the Reynolds number past a turbulent threshold \cite[]{lathrop92,cadot97}.

\subsection{\label{sec:bifglob}Higher-Reynolds number: multistability and turbulent bifurcation}

From all the above reported observations in the negative direction of rotation, we conclude that the transition to turbulence is completed at $Re_t$ and that the azimuthal kinetic energy fluctuation can be considered as an order parameter for the global transition, from the onset of time-dependence $Re_c=330$ to the fully turbulent state transition/crossover at $Re_t=3.3 \times 10^3$, \ie over a decade in Reynolds number. 

In the inertial regime above $Re_t$, the von K\'arm\'an flow driven by high-curvature bladed impellers rotating in the negative direction presents another original behaviour:  \citet{ravelet2004} have shown that the turbulent von K\'arm\'an flow can exhibit multistability at high-Reynolds number. To study and analyse this phenomenon, it is necessary to introduce an additional parameter with respect to the present paper study: the rotation velocity difference $\Delta f=f_2-f_1$ between the two impellers. The so-called \lq\lq Turbulent bifurcation\rq\rq ~and multistability are observed exclusively for the negative direction of rotation. So, the $\Delta f=0$ regime presented along this paper ---called (s) for symmetric in \citet{ravelet2004}--- can be observed only if both motors are started together, \ie if $\Delta f$ is kept equal to zero at any time. Once some velocity difference is applied long enough ---depending of the magnitude of $|\Delta f|$---, the flow changes abruptly to a one cell flow with axial pumping towards one of the impellers only instead of towards each impeller. This new flow ---called (b) for bifurcated in \citet{ravelet2004}--- strongly breaks the \zob-symmetry, has no middle shear-layer and requires much higher torque from the motors: typically 3 times the value of (s)-flow, with a finite difference between the two motors. The mean reduced torque at $\Delta f = 0$ is plotted with stars in figure~\ref{fig:kpdere_anti}: branches (s) and (b) co-exist for $Re \gtrsim Re_m=10^4$. To recover the \zob-symmetric flow, one should stop the motors or at least decrease $Re$ below $Re_m$.

It is worth noting that this multistability is only observed above $Re_t$, \ie for flows with a well developed turbulent inertial Kolmogorov cascade. Furthermore, cycles in the parameter plane $\{K_{p2}-K_{p1} ; f_2-f_1\}$ have been made for various $Re$ between $100$ and $3 \times 10^5$ \cite[]{theseflo}. At low-Reynolds numbers ---$Re \lesssim 800$---, this cycle is reduced to a continuous, monotonic and reversible line in the parameter plane. The first apparition of \lq\lq topological\rq\rq~ transformations of this simple line into multiples discontinuous branches of a more complex cycle is reported at $Re \simeq 5\times 10^3$, in the neighbourhood of the transitional Reynolds number $Re_t$, and multistability for $\Delta f = 0$ is first observed for $Re \sim 10^4$. The extensive study of this turbulent bifurcation with varying $Re$ is worth a complete article and will be reported elsewhere. 

From the above preliminary report of our results, we emphasize the fact that the turbulent bifurcation seems really specific of fully developed turbulent flows. Whereas the exact counter-rotating flow ($\Delta f=0$) will never bifurcate \cite[]{ravelet2004}, for a small $\Delta f $ ($0 < |\Delta f|/f \ll 1$) this turbulent bifurcation around $Re_m = 10^4$ will correspond to a first order transition on the way to infinite Reynolds number dynamics: this flow really appears as an ideal prototype of an ideal system undergoing a succession of well-defined transitions on the way from order to high-Reynolds-number turbulence.

\subsection{The regimes: a summary}
 \label{sec:summary}

The next section concerns some aspects specific to the inertial stirring. Thereafter is the discussion (\S~\ref{sec:discussion}) about the role of the symmetries and of the spatial scales of the flow which can be read almost independently. The following summary of the observed regimes and transitions is given as a support for the discussion.

\begin{itemize}

\item $Re<175$ : $m=0$, axisymmetric, \zob-symmetric steady basic flow (\S~\ref{sec:basic}),
\item $175<Re<330$ : $m=2$, discretely \zob-symmetric steady flow (\S~\ref{sec:m=2}),
\item $330<Re<389$ : $m=2$, non \zob-symmetric, equatorial-parity-broken travelling waves (\S~\ref{sec:pres:TW}, \S~\ref{sec:quant:330}),
\item $389<Re<400$ : modulated travelling waves (\S~\ref{sec:pres:MTW}),
\item $400<Re<408$ : chaotic modulated travelling waves (\S~\ref{sec:pres:MTW}),
\item $400<Re \lesssim 1000$ : chaotic flow (\S~\ref{sec:pres:chaos}, \S~\ref{sec:quant:400}),
\item $1000 \lesssim Re \lesssim 3300$ : transition to turbulence (\S~\ref{sec:pres:high_freq_spectr}, \S~\ref{sec:quant:1000}),
\item $Re \gtrsim 3300$ : inertially-driven fully turbulent flow (\S~\ref{sec:quant:3300}),
\item $Re \gtrsim 10^4$ : multivalued inertial turbulence regimes (\S~\ref{sec:bifglob}).

\end{itemize}

\section{Viscous stirring vs. inertial stirring}
\label{sec:visc2inert}

\begin{figure*}
\begin{center}
\includegraphics[clip,width=120mm]{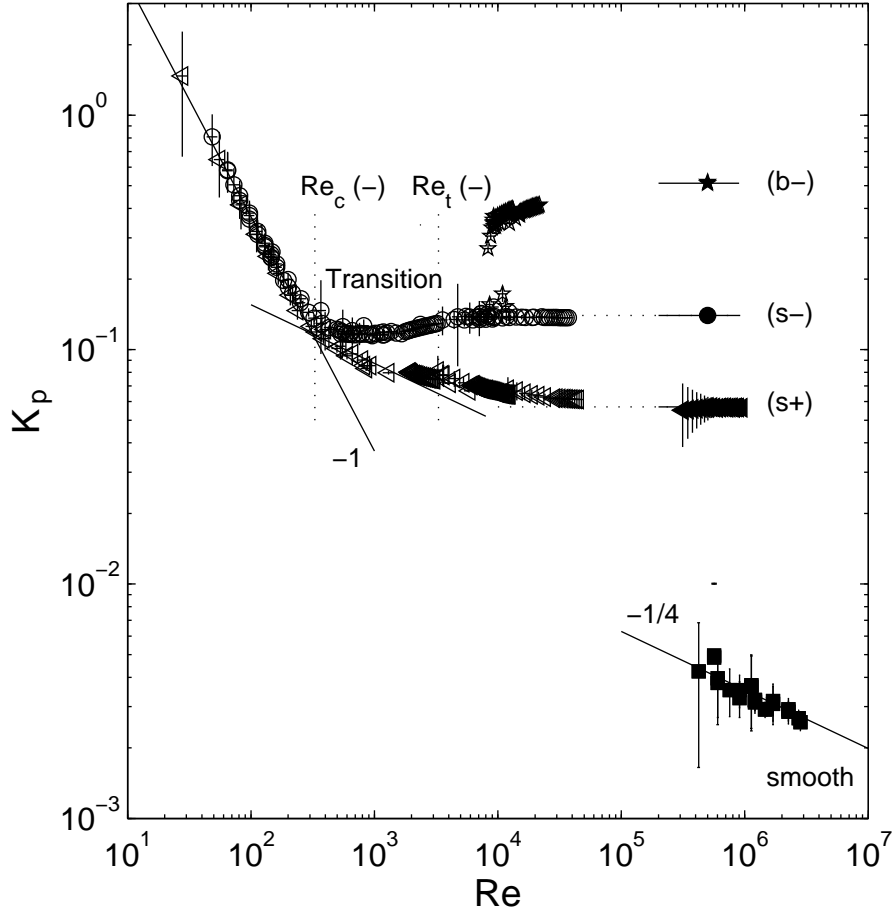}
\caption{Compilation of the dimensionless torque $K_p$ \vs $Re$ for various flows. All data stands for \zob -symmetric von-K\'arm\'an flows except the branch labelled (b$-$) ($\star$): see caption of figure~\ref{fig:kpdere_anti} for details. 
($\circ$)~: direction of rotation $(-)$. ($\triangleleft$)~: direction of rotation $(+)$; the solid line is a non-linear fit of equation $K_p=36.9\times Re^{-1}$ between $Re=30$ and $Re=250$. 
Some data for flat disks of standard machine shop roughness, operated in pure water up to $25$Hz (squares) are also displayed with a $Re^{-1/4}$ fit. 
Another $-1/4$ power law is fitted for the positive direction of rotation for $330 \leq Re \leq 1500$ and is displayed between $Re=10^2$ and $Re=10^4$.
Relative error on $Re$ is $\pm10\%$~; absolute error of $\pm0.1 \ {\rm N.m}$ on the torque. 
$Re_c$ and $Re_t$ are the transition values computed from the fits of figure~\ref{fig:niveaubruitdere}.}
\label{fig:kpdere_complete}
\end{center}
\end{figure*}

We now focus on the specificities of the inertial stirring. In the preceding parts, a single rotation sense, the negative $(-)$, was studied. However, very relevant information can be obtained from the comparison of data in both senses of impellers rotation, which is equivalent to have two sets of impellers with opposite curvature at any time in the same experiment.

The guideline for this analysis is the global energetic measurements along the whole Reynolds number range. The data for sense $(-)$ have already been partly discussed in the preceding part (figure~\ref{fig:kpdere_anti}), but the full set comes here in figure~\ref{fig:kpdere_complete}. At low Reynolds number the two curves are identical, which means that the blades have no effect on the viscous stirring. This is analysed in \S~\ref{subsec:visc2inert}. However, at high Reynolds number, there is a factor 3 between both curves, denoting very different inertial regimes, as discussed in \S~\ref{subsec:inertial}. 

\subsection{From viscous to inertial stirring}
\label{subsec:visc2inert}

While $Re \lesssim 300$, the dimensionless torque $K_p$ scales as $Re^{-1}$. We are in the laminar regime \cite[]{schlichting} and the viscous terms are dominant in the momentum balance.
These regimes correspond to $m=0$ or $m=2$ steady flows, with an eventual slow drift (\S~\ref{sec:basic} \& \ref{sec:m=2}).

From the power consumption point of view, both directions of rotations are equivalent. The two curves ---circles for direction $(-)$ and left triangles for direction $(+)$--- collapse for $Re\lesssim 300$ on a single curve of equation $K_p=36.9 Re^{-1}$.

\begin{figure}
\begin{center}
\includegraphics[clip,width=80mm]{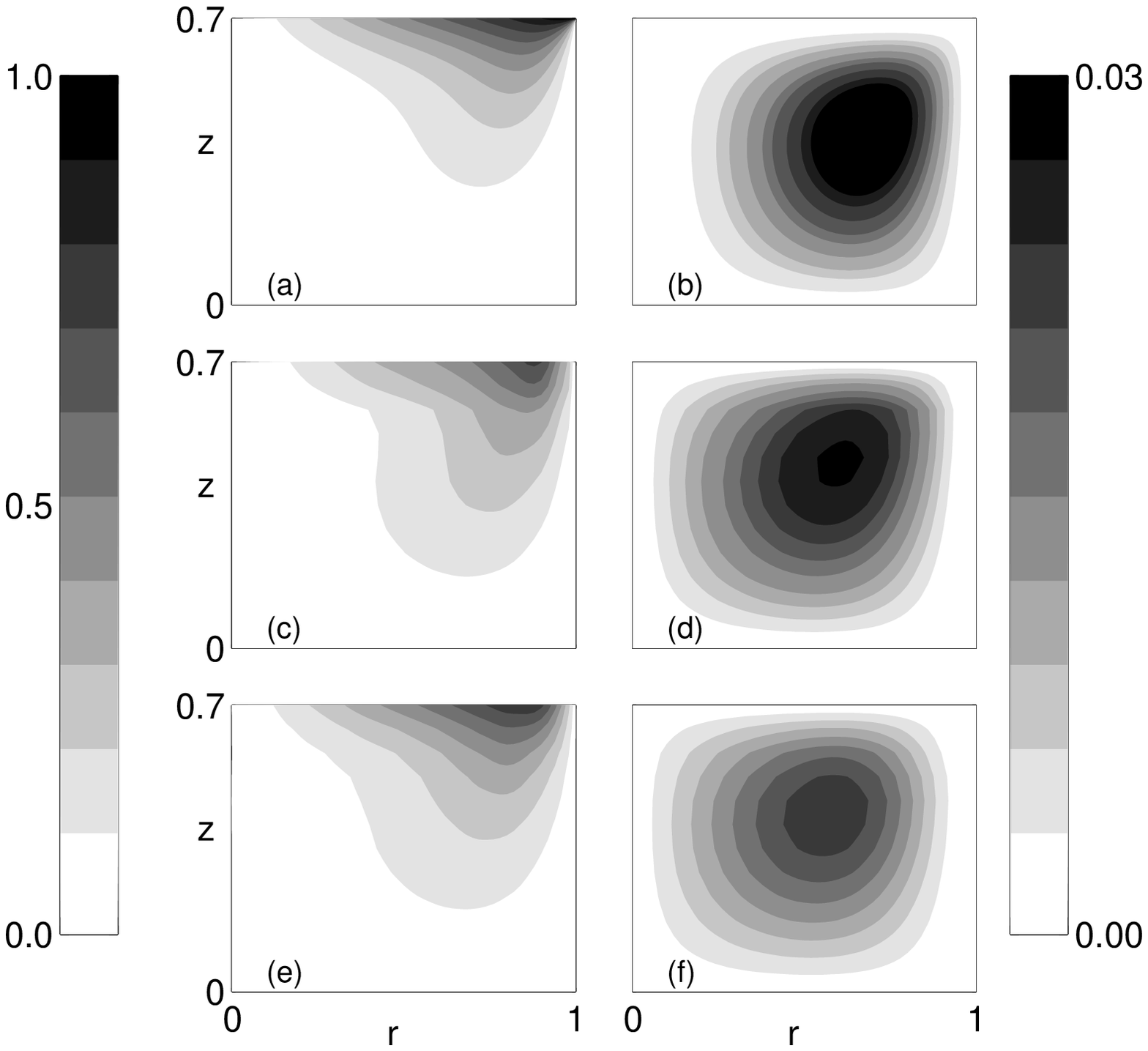}
\caption{Comparison between a numerical simulation (a-b) performed with the code of \citet{nore03} in a cylinder of aspect ratio $\Gamma=1.4$ at $Re=120$ and two experimental velocity fields measured by LDV in direction $(+)$ at $Re=130$ (c-d) and in direction $(-)$ at $Re=120$ (e-f). The flow quantities which we present are in (a-c-e) the azimuthal velocity $v_{\theta}$ and in (b-d-f) the poloidal stream function $\Psi$. Presenting the fields between $0 \leq r \leq 1$ and $0 \leq z \leq 0.7$ is sufficient due to axisymmetry and \zob -symmetry. Blades or smooth disk are at $z=0.7$.}
\label{fig:comparaisoncaro}
\end{center}
\end{figure}

We performed velocity field measurements for the two flows at $Re \simeq 120-130$ (figure~\ref{fig:comparaisoncaro}c-f). The differences between the two directions are minor. The order of magnitude of the mean poloidal and toroidal velocities are the same within $15\%$ for both directions of rotation in the laminar regime, whereas at very high $Re$, they strongly differ (by a factor $2$) \cite[]{ravelet2005}.

The flow is thus not sensitive to the shape of the impeller blades in the laminar regime. To explain this, we make the hypothesis that for these large impellers of radius $0.925 R_c$, fitted with blades of height $h=0.2 R_c$, the flow at low $Re$ is equivalent to the flow between flat disks in an effective aspect ratio $\Gamma=(H-2h)/R_c=1.4$. \citet{nore04jfm} numerically studied the flow between counter-rotating smooth flat disks enclosed in a cylinder and report the dependence of the first unstable mode wave number on the aspect ratio $\Gamma=H/R_c$. In their computations, the critical wave number is $m=1$ for $\Gamma=1.8$, whereas for $\Gamma=1.4$, it is $m=2$ as we do observe in our experiments.

We thus compare in figure~\ref{fig:comparaisoncaro} our experimental velocity fields to a numerical simulation performed by Caroline Nore at the same $Re$ and in aspect ratio $\Gamma=1.4$. The three fields are very close. A possible physical explanation for this effect is the presence of viscous boundary layers along the resting cylinder wall. The typical length scale of the boundary layer thickness can be estimated as $\delta=Re^{-1/2}$. At the Reynolds number when the impellers blades start to become visible, \ie at $Re \simeq 300$, this boundary layer thickness is of the order of $\delta \simeq 6 \ {\rm mm}$, while the gap between the impellers and the cylinder wall is $7.5 \ {\rm mm}$. It is also of the order of magnitude of the minimum distance between two blades. For $Re \lesssim 300$, the fluid is thus kept between the blades and can not be expelled radially: it rotates solidly with the impellers. The stirring cannot be considered as inertial and does not depend on the blades shape. 

For $Re \gtrsim 300$, the dimensionless torque starts to shift from a $Re^{-1}$ law and simultaneously discriminates between both sense of rotation: the inertial stirring becomes dominant over the viscous stirring. Simultaneously also, the steady flow becomes unstable with respect to time-dependence (\S~\ref{sec:descr_regimes} and \ref{sec:quant}
).

\subsection{Inertial effects}

\label{subsec:inertial}

At high Reynolds number, we observe in figure~\ref{fig:kpdere_complete} different behaviours for $K_p$ in both sense of rotation. Sense $(-)$ passes a minimum for $Re \simeq 1000$ and then rapidly reaches a flat plateau above $Re_t=3300$ (see \S~\ref{sec:fullturb}), whereas sense $(+)$ asymptotically reaches a regime with only a third of the power dissipation of sense $(-)$. Together with the curved blade data, figure~\ref{fig:kpdere_complete} presents additional data for smooth disks. The dimensionless torque $K_p$ is approximately $30$ times smaller for smooth disks than for bladed disks, and does not display any plateau at high-Reynolds number but a $Re^{-1/4}$ scaling law, as described by \citet{cadot97}. 

It is tempting to compare our curve $K_p(Re)$ with the classical work of \citet{nikuradse32,nikuradse33} consisting in a complete and careful experimental data set about the turbulence in a pipe flow with controlled wall roughness. The data concern the friction factor ---equivalent of $K_p$--- measured over a wide range between $Re=500$ and $Re=10^6$, which is shown to strongly depend of the wall roughness above $Re \simeq 3000$. The wall roughness is made by controlled sand grains of diameter in the range $1/507$ to $1/15$ of the pipe radius, somewhat smaller than our blades height $h/R_c = 1/5$ which can be thought as an effective roughness. 

This data set has defied theory along decades and still motivates papers. Recently, \citet{goldenfeld2006} and \citet{gioia2006} proposed phenomenological interpretations and empirical reduction of Nikuradse's data. In few words, both recent works connect the very high-Reynolds inertial behaviour ---a plateau at a value which scales with the roughness to the power 1/3--- to the Blasius $Re^{-1/4}$ law for the dissipative region at intermediate $Re$. \citet{goldenfeld2006}, using a method from critical point physics, finds a scaling for the whole domain above $Re \simeq 3000$, whereas \citet{gioia2006} describe the friction factor over the same Reynolds range according to Kolmogorov's phenomenological model.

Compared with pipe flow results and models, our $K_p(Re)$-curve (figure~\ref{fig:kpdere_complete}) looks very similar except for the region \citet{gioia2006} called the energetic regime. Indeed, in our specific case the basic flow itself is already dominated by vortices of the size of the vessel. The negative direction (circles in figure~\ref{fig:kpdere_complete}) shows a minimum followed by a plateau above $Re_t =3300$ and is in agreement with the general inertial behaviour described above. However for the positive direction (left triangles in figure~\ref{fig:kpdere_complete}), the $K_p$ curve seems continuously decreasing up to $Re \simeq 10^6$. Looking closer, one can observe a short $Re^{-1/4}$ Blasius regime for $Re$ between $300$ and $1500$ ---highlighted by a fit in figure~\ref{fig:kpdere_complete}--- followed by a very slow variation over the next two decades: logarithmic corrections are still visible in the range $10^4 \lesssim Re \lesssim 5 \times 10^4$. For this direction it is more difficult to define a threshold for the plateau observed in pure water \cite[]{theselouis}. Nevertheless, this threshold should be of order of $10^5$, \ie much higher than with negative rotation. 

A possible explanation of this strong difference may rely in the structure of the flow inside the impellers, \ie in-between the blades. Let us first assume that this flow is dominated by what happens along the extrados of the blades, on which the pressure is the higher. Then we can assume that the blades curvature leads to stable boundary layers in positive rotation and to Goertler instability in negative rotation. The first case develops Blasius boundary layers, whereas the latter develops turbulent boundary layers with much more vortices. Therefore, when the boundary layer detaches ---somewhere along the blades or at least at their end--- the Blasius boundary layer in the positive rotation sheds less turbulent vortices than the Goertler's unstable layer does in the negative rotation.

The above description can be sufficient to explain why the negative rotation is able to produce a Kolmogorov cascade even at quite low-Reynolds numbers near $Re_t$. However if, in the positive rotation case, the flow is only seeded by vortices produced by the stable boundary layer which develops along the smooth blade faces, it is clear that a Blasius $Re^{-1/4}$ can be observed in this transition Reynolds range and that a full inertial regime does not occur below a very high-Reynolds number owing to the very small roughness of the blades faces. This could be why both curves in figure~\ref{fig:kpdere_complete} look so different: the lower one looks qualitatively like a low-roughness boundary flow and the upper one looks like a high-roughness boundary flow. Anyway, this may only account for a part of the flow driving: the resistive torque is much higher for any bladed impellers than for flat disks as shown in figure~\ref{fig:kpdere_complete}.

Our observation of the closed von K\'arm\'an turbulent flow is thus consistent with the claim by \citet{goldenfeld2006} that full understanding of turbulence requires explicit accounting for boundary roughness.

\section{Discussion and conclusion}
\label{sec:discussion}

\subsection{Symmetries and first bifurcations}
\label{sec:disc:symmetries}

As for many flows, the similarity of the flow behaviour at low Reynolds number with intermediate-size non-linear system is obvious: breaking a spatial symmetry first, then a temporal symmetry and finally transit to chaos by a quasi-periodic scenario.

Comparable study has been carried both experimentally and numerically in the von K\'arm\'an flow with flat disk and variable aspect ratio by \citet{nore03,nore04jfm,nore05}. Our results agree well with their results on the first instability mode $m=2$ if considering the fluid in the blade region as almost solidly driven, which reduces the aspect ratio (see \S~\ref{subsec:visc2inert}). However, all thresholds appear at much lower $Re$ for bladed impellers than for flat disks: $175$ \vs $300$ for the first steady bifurcation and $330$ \vs above $600$ for the first temporal instability of $m=2$ mode, not observed in \citet{nore05} study.

Another important difference between both system concerns its symmetries. Whereas Nore and collaborators deal with exact counter rotation by using a single motor to drive both disks, our experimental setup uses two independent motors and reaches only a approximation on a counter-rotating regime. As a consequence, the \zob -symmetry is \emph{stricto sensu} broken at any Reynolds number and the group of symmetry of our problem is $SO(2)$ instead of $O(2)$. To evaluate the level of symmetry breaking we can use a small parameter \cite[]{chossat93,porter2005}, \emph{e.g.} $\epsilon=(f_1-f_2)/(f_1+f_2)$ which is between $10^{-4}$ and $10^{-3}$ in our runs. 

Carefully controlling this parameter is an interesting issue: recently, in almost the same von K\'arm\'an flow in the positive sense of rotation at high $Re$, \citet{torre2007} reported bistability and a turbulent bifurcation at exactly $\epsilon=0$ between two \zob-symmetric flows. For non-zero $\epsilon$, the mixing layer lies slightly above or below the equator and it randomly jumps between these two symmetric positions when $\epsilon$ is carefully set to zero.

With our very small experimental $\epsilon$, we verify theoretical predictions \cite[]{chossat93,porter2005} for the 1:2 spatial resonance or $k-2k$ interaction mechanism with slightly broken reflexion symmetry. Instead of mixed mode, pure mode and heteroclinic cycles ---specific of $O(2)$ and carefully reported by \citet{nore03,nore04jfm,nore05}--- we only observe drifting instability patterns, \ie travelling waves and modulated travelling waves, characteristic of $SO(2)$. Also, the drift frequency is very close to zero at the threshold $Re_c=330$ (figure~\ref{fig:QP}a), in agreement with the prediction $f_D \sim \cal{O}(\epsilon)$ \cite[]{chossat93,porter2005}. This bifurcation to travelling waves is similar to the 1-D drift instability of steady patterns, observed in many systems \cite[see, \emph{e.g.}][]{fauve91}. It relies on the breaking of the parity ($\theta \rightarrow -\theta$) of the pattern \cite[]{coullet90}: the travelling-wave pattern is a pair of tilted vortices. The bifurcation is an imperfect pitchfork \cite[]{porter2005}.

Finally, the comparison can be extended to the travelling waves observed with flat disks above the mixed and pure modes \cite[]{nore03,nore05}. The observed wave frequencies are of the same order of magnitude in both case, which let us believe that the same kind of hydrodynamics is involved , \ie the blades play again a minor role at these low Reynolds numbers. However, the frequency ratio between the basic waves (TW) and their modulations (MTW) at onset is much higher ($\sim32$) in our experiment than in the numerical simulations ($\sim5$) \cite[]{nore03}. This could be  due to the high number of blades.

We also wish to consider the symmetry of the von K\'arm\'an flow with respect to the rotation axis. In fact, the time-averaged flow is exactly axisymmetric while the instantaneous flow is not, because of the presence of blades. However, axisymmetry can be considered as an effective property at any time at low-Reynolds number and at least up to $Re=175$, since we have shown that the blades have almost no effect on the flow (see \S~\ref{subsec:visc2inert}). With increasing $Re$, the blades start playing their role and effectively break the axisymmetry of the instantaneous flow.

Finally, we emphasize that the observations made below $Re \sim 400$ closely remind the route to chaos trough successive symmetry break for low degree of freedom dynamical systems. Our system can thus be considered as a small system ---in fact this is coherent with the aspect ratio which is of order of $1$--- until the Reynolds number becomes high enough to excite small dynamical scales in the flow.

\subsection{The three scales of the von K\'arm\'an flow}
\label{sec:disc:3scales}

The observations reported in this article ---visualisations, spectra--- evidenced three different scales. In particular, time-spectra contain two time-frequency domains above and below the injection frequency $f_a=f$. Let us first make a rough sketch of the correspondence between temporal and spatial frequency scales of the whole flow:

\begin{itemize}

\item the smallest space-frequencies, at the scale of the vessel, describe the basic swirling flow due to the impeller and produce the intermediate frequency-range, \ie the peak at $f_a=f$ in the time-spectrum;
\item the intermediate space-frequencies due to the shear-layer main instabilities produce the lowest time-frequencies;
\item the highest space-frequencies produce, of course, the highest temporal frequencies, \ie the Kolmogorov region. 

\end{itemize}

The Taylor's hypothesis is based on a linear mapping between space- and time-frequencies. It is probably valid for the high part of the spectrum, but the mapping might be not linear and even not monotonic for the low part. We discuss each part of the spectrum in the two following paragraphs.

\subsubsection{The $1/f$ low-frequency spectrum}

Once chaos is reached at $Re = 400$, a strong continuous and monotonic low-frequency spectrum is generated (Fig.~\ref{fig:ldvprecision1}h). In the chaotic regime below $Re \sim 1000$, the spectrum evolves to a neat $-1$ power law. Then, this part of the spectrum does not evolve any more with $Re$.

Low-frequency $-1$ exponents in spectra are common and could be due to a variety of physical phenomena: so-called \lq\lq$1/f$ noises\rq\rq~ have been widely studied,{\em e.g.}, in the condensed matter field \cite[see for instance][]{dutta1981}.

For turbulent von K\'arm\'an flows driven by two counter-rotating impellers, this low time-scale dynamics has been already observed over at least a decade in liquid helium by \citet{zocchi94} as well as for the magnetic induction spectrum in liquid metals \cite[]{bourgoin02,volk2006pof}. However, experiments carried on a one-cell flow ---without turbulent mixing-layer--- did not show this behaviour \cite[]{theselouis,ravelet2004,theseflo}. We therefore conclude that the $1/f_a$-spectrum is related to the chaotic wandering of the mixing-layer which statistically restores the axisymmetry. Once again, the mixing-layer slow dynamics dominates the whole dynamics of our system, from momentum transfer \cite[]{marie04pof} to the very high level of turbulent fluctuations (Fig.~\ref{fig:niveaubruitdere} and \ref{fig:PDFchameau}).

Furthermore, we can make the hypothesis that the $-1$ slope is due to the distribution of persistence times in each side of the bimodal distribution (Fig.~\ref{fig:PDFchameau}): the low-frequency part of the spectrum can be reproduced by a random binary signal. Similar ideas for the low-frequency spectral construction are proposed for the magnetic induction in the von K\'arm\'an sodium (VKS) experiment \cite[]{ravelet2007}. In both cases, longer statistics would be needed to confirm this idea.

\subsubsection{The turbulent fluctuations}
\label{subsubsec:taylor}

We above emphasize how the flow transits from chaos to turbulence between $Re \simeq 1000$ and $Re_t = 3300$. We label this region \lq\lq transition to turbulence\rq\rq ~and observe the growth of a power-law region in the time-spectra for $f_a > f$. Does this slope trace back the Kolmogorov cascade in the space-spectra? 

As the classical Taylor hypothesis cannot apply to our full range spectrum, we follow the Local Taylor Hypothesis idea \cite[]{pinton94} for the high-frequency part $f_a>f$. 
Whereas \citet{pinton94} did not apply their technique ---using instantaneous velocity instead of a constant advection--- to the extreme case of zero advection, we think it can be applied here owing to the shape of the azimuthal velocity PDF (figure~\ref{fig:PDFchameau}). These distributions show first that the instantaneous zero velocity is a quite rare event: a local minimum of the curve. The modulus of velocity spends typically $75\%$ of the time between $1/2 V_m$ and $3/2 V_m$, where $\pm V_m$ are the positions of the PDF maxima. The sign of the advection has no effect on the reconstructed wave number. We can thus conclude that frequency and wave number modulus can be matched each other at first order by a factor equal to the most probable velocity $|V_m|$ or by the mean of $|v_{\theta}|$, both very close to each other. 
This approach is coherent with a binary view of the local turbulent signal jumping randomly between two opposite mean values, just as in turbulent flow reversal model of, e.g., \citet{benzi2005}. Then, the high-frequency part of the spectrum is equivalent to the spectrum obtained by averaging the spectra of every single time-series between jumps, while the low-frequency part is dominated by dynamics of the jumps themselves.

Owing to these arguments, we are convinced that an algebraic region dominates the high-frequency part of $k$-spectra above $Re_t$. Observed exponents ($-1.55$) are of the order of the Kolmogorov exponent, less than $10\%$ smaller in absolute value. Similar exponents are also encountered at other locations in the vessel. 

\subsection{\label{sec:conclusion}Conclusion and perspectives}

The von K\'arm\'an shear-flow with inertial stirring has been used for a global study of the transition from order to turbulence. The transition scenario is consistent with a globally supercritical scenario and this system appears as a very powerful table-top prototype for such type of study. We have chosen to emphasize a global view over a wide range of Reynolds number. This allowed to make connections between informations relaying on local (velocities) or global quantities (torques, flow symmetries), as discussed in \S~\ref{subsec:inertial} and \ref{sec:discussion}.

\subsubsection{Going further}
As a perspective, it would be first interesting to increase the resolution of the analysis next to the different observed thresholds. It would also be worthwhile to perform the same wide-range study for the other sense of rotation $(+)$ or another couple of impellers. Finally, these studies would enable a comparison of the inertial effects on the turbulent dynamics at very high Reynolds number. 

\subsubsection{Controlling the mixing layer}
Many results of the present study proceed from velocity data collected in the middle of the shear-layer and we have shown that this layer and its chaotic/turbulent wandering can be responsible for the low frequency content of the chaotic/turbulent spectrum of the data.

With the slightly different point of view of controlling the disorder level, we have modified the dynamics of the shear-layer by adding a thin annulus located in the mid-plane of the flow \cite[]{ravelet2005}. This property was recently used in the Von K\'arm\'an Sodium (VKS) experiment held at Cadarache, France and devoted to the experimental study of dynamo action in a turbulent liquid sodium flow. Dynamo has effectively been observed for the first time in this system with a von K\'arm\'an configuration using, among other characteristics, an annulus in the mid-plane \cite[]{monchaux2006b} and is sensitive to the presence of this device. Moreover, clear evidence has been made that the mixing-layer large-scale patterns have a strong effect on the magnetic field induction at low frequency \cite[]{volk2006pof, ravelet2007}. Further studies of this effect in water experiments are under progress.

\subsubsection{Statistical properties of the turbulence}
Studies of the von K\'arm\'an flow currently in progress invoke both a wider range of data in space, with the use of Stereoscopic 3-components Particle Image Velocimetry (SPIV) and a wider range in Reynolds number. 

Whereas the SPIV is slower than LDV and will not allow time-spectral analysis, it offers a global view of the flow and allows to characterize statistical properties of the turbulent velocity. Guided by the behaviour of the variance of the local azimuthal velocity revealed in the present article (figure~\ref{fig:niveaubruitdere}), we expect to analyse the evolution of the spatio-temporal statistical properties with $Re$. Such study is very stimulating for theoretical advances toward a
statistical mechanics of the turbulence in 2D \cite[]{sommeria1991,chavanis1998}, quasi 2D \cite[]{bouchet2002,jung2006} or
axisymmetric flows \cite[]{leprovost2006,monchaux2006a}.

\begin{acknowledgments}
We are particularly indebted to Vincent Padilla and C\'ecile Gasquet for building up and piloting the experiment. We acknowledge Caroline Nore for making her simulations available, Arnaud Guet for his help on the visualisations and Fr\'ed\'eric Da Cruz for the viscosity measurements. We have benefited of very fruitful discussions with B. Dubrulle, N. Leprovost, L. Mari\'e, R. Monchaux, C. Nore, J.-F. Pinton and R. Volk.
\end{acknowledgments}

\bibliographystyle{plainnat}
\bibliography{BiblioVKTurb}

\begin{thebibliography}{66}
\expandafter\ifx\csname natexlab\endcsname\relax\def\natexlab#1{#1}\fi
\expandafter\ifx\csname url\endcsname\relax
  \def\url#1{{\tt #1}}\fi

\bibitem[Batchelor(1951)]{batchelor51}
G.~K. Batchelor.
\newblock Note on a class of solutions of the {N}avier-{S}tokes equations
  representing steady rotationally-symmetric flow.
\newblock {\em Q. J. Mech. App. Math.}, 4:\penalty0 29, 1951.

\bibitem[Benzi(2005)]{benzi2005}
R.~Benzi.
\newblock Flow reversal in a simple dynamical model of turbulence.
\newblock {\em Phys. Rev. Lett.}, 95:\penalty0 024502, 2005.

\bibitem[Bouchet and Sommeria(2002)]{bouchet2002}
F.~Bouchet and J.~Sommeria.
\newblock Emergence of intense jets and {J}upiter's great red spot as
  maximum-entropy structures.
\newblock {\em J. Fluid Mech.}, 464:\penalty0 165--207, 2002.

\bibitem[Bourgoin et~al.(2002)Bourgoin, Mari{\'e}, P{\'e}tr{\'e}lis, Gasquet,
  Guigon, Luciani, Moulin, Namer, Burguete, Chiffaudel, Daviaud, Fauve, Odier,
  and Pinton]{bourgoin02}
M.~Bourgoin, L.~Mari{\'e}, F.~P{\'e}tr{\'e}lis, C.~Gasquet, A.~Guigon, J.-B.
  Luciani, M.~Moulin, F.~Namer, J.~Burguete, A.~Chiffaudel, F.~Daviaud,
  S.~Fauve, P.~Odier, and J.-F. Pinton.
\newblock {MHD} measurements in the von {K}\'arm\'an sodium experiment.
\newblock {\em Phys. Fluids}, 14:\penalty0 3046, 2002.

\bibitem[Buchhave et~al.(1979)Buchhave, George, and Lumley]{buchhave79}
P.~Buchhave, W.~K. George, and J.~L. Lumley.
\newblock The measurement of turbulence with the {L}aser-{D}oppler anemometer.
\newblock {\em Annu. Rev. Fluid Mech.}, 11:\penalty0 443, 1979.

\bibitem[Cadot et~al.(1997)Cadot, Couder, Daerr, Douady, and Tsinober]{cadot97}
O.~Cadot, Y.~Couder, A.~Daerr, S.~Douady, and A.~Tsinober.
\newblock Energy injection in closed turbulent flows: Stirring through boundary
  layers versus inertial stirring.
\newblock {\em Phys. Rev. E}, 56:\penalty0 427, 1997.

\bibitem[Cadot et~al.(1995)Cadot, Douady, and Couder]{cadot95}
O.~Cadot, S.~Douady, and Y.~Couder.
\newblock Characterization of the low-pressure filaments in a three-dimensional
  turbulent shear flow.
\newblock {\em Phys. Fluids}, 7:\penalty0 630, 1995.

\bibitem[Chavanis and Sommeria(1998)]{chavanis1998}
P.-H. Chavanis and J.~Sommeria.
\newblock Classification of robust isolated vortices in two-dimensional
  hydrodynamics.
\newblock {\em J. Fluid Mech.}, 356:\penalty0 259--296, 1998.

\bibitem[Chossat(1993)]{chossat93}
P.~Chossat.
\newblock Forced reflexional symmetry breaking of an {\em{o}(2)}-symmetric
  homoclinic cycle.
\newblock {\em Nonlinearity}, 6:\penalty0 723--731, 1993.

\bibitem[Coullet and Iooss(1990)]{coullet90}
P.~Coullet and G.~Iooss.
\newblock Instabilities of one-dimensional cellular patterns.
\newblock {\em Phys. Rev. Lett.}, 64:\penalty0 866, 1990.

\bibitem[de~la Torre and Burguete(2007)]{torre2007}
A.~de~la Torre and J.~Burguete.
\newblock Slow dynamics in a turbulent von {K}\'arm\'an swirling flow.
\newblock {\em Phys. Rev. Lett.}, 99:\penalty0 054101, 2007.

\bibitem[Douady et~al.(1991)Douady, Couder, and Brachet]{douady91}
S.~Douady, Y.~Couder, and M.~E. Brachet.
\newblock Direct observation of the intermittency of intense vorticity
  filaments in turbulence.
\newblock {\em Phys. Rev. Lett.}, 67:\penalty0 983, 1991.

\bibitem[Dutta and Horn(1981)]{dutta1981}
P.~Dutta and P.~M. Horn.
\newblock Low-frequency fluctuations in solid: 1/f noise.
\newblock {\em Rev. Mod. Phys.}, 53:\penalty0 497, 1981.

\bibitem[Escudier(1984)]{escudier84}
M.~P. Escudier.
\newblock Observations of the flow produced in a cylindrical container by a
  rotating endwall.
\newblock {\em Exp. Fluids}, 2:\penalty0 189, 1984.

\bibitem[Falkovich(1994)]{falkovich1994}
G.~Falkovich.
\newblock Bottleneck phenomenon in developed turbulence.
\newblock {\em Phys. Fluids}, 6:\penalty0 1411, 1994.

\bibitem[Fauve et~al.(1991)Fauve, Douady, and Thual]{fauve91}
S.~Fauve, S.~Douady, and O.~Thual.
\newblock Drift instabilities of cellular patterns.
\newblock {\em J. Phys. II}, 1:\penalty0 311--322, 1991.

\bibitem[Fauve et~al.(1993)Fauve, Laroche, and Castaing]{fauve93}
S.~Fauve, C.~Laroche, and B.~Castaing.
\newblock Pressure fluctuations in swirling turbulent flows.
\newblock {\em J. Phys. II}, 3:\penalty0 271, 1993.

\bibitem[Frisch(1995)]{frisch95}
U.~Frisch.
\newblock {\em Turbulence - {T}he legacy of {A}. {N}. {K}olmogorov}.
\newblock Cambridge University Press, New-York, 1995.

\bibitem[Gauthier et~al.(1999)Gauthier, Gondret, and Rabaud]{gauthier99}
G.~Gauthier, P.~Gondret, and M.~Rabaud.
\newblock Axisymmetric propagating vortices in the flow between a stationary
  and a rotating disk enclosed by a cylinder.
\newblock {\em J. Fluid Mech.}, 386:\penalty0 105, 1999.

\bibitem[Gelfgat et~al.(1996)Gelfgat, Bar-Yoseph, and Solan]{gelfgat1996}
A.~Y. Gelfgat, P.~Z. Bar-Yoseph, and A.~Solan.
\newblock Steady states and oscillatory instability of swirling flow in a
  cylinder with rotating top and bottom.
\newblock {\em Phys. Fluids}, 8:\penalty0 2614, 1996.

\bibitem[Gioia and Chakraborty(2006)]{gioia2006}
G.~Gioia and P.~Chakraborty.
\newblock Turbulent friction in rough pipes and the energy spectrum of the
  phenomenological theory.
\newblock {\em Phys. Rev. Lett.}, 96:\penalty0 044502, 2006.

\bibitem[Goldenfeld(2006)]{goldenfeld2006}
N.~Goldenfeld.
\newblock Roughness-induced critical phenomena in a turbulent flow.
\newblock {\em Phys. Rev. Lett.}, 96:\penalty0 044503, 2006.

\bibitem[Harriott and Brown(1984)]{harriott84}
G.~M. Harriott and R.~A. Brown.
\newblock Flow in a differentially rotated cylindrical drop at moderate
  {R}eynolds number.
\newblock {\em J. Fluid Mech.}, 144:\penalty0 403, 1984.

\bibitem[Hodgman(1947)]{handbook}
C.~D. Hodgman, editor.
\newblock {\em Handbook of chemistry and physics{,} thirtieth edition}.
\newblock Chemical rubber publishing co., 1947.

\bibitem[Jung et~al.(2006)Jung, Morrison, and Swinney]{jung2006}
S.~W. Jung, P.~J. Morrison, and H.~L. Swinney.
\newblock Statistical mechanics of two-dimensional turbulence.
\newblock {\em J. Fluid Mech.}, 554:\penalty0 433--456, 2006.

\bibitem[Knobloch(1996)]{knobloch96}
E.~Knobloch.
\newblock Symmetry and instability in rotating hydrodynamic and
  magnetohydrodynamic flows.
\newblock {\em Phys. Fluids}, 8:\penalty0 1446, 1996.

\bibitem[{La Porta} et~al.(2001){La Porta}, Voth, Crawford, Alexander, and
  Bodenschatz]{laporta01}
A.~{La Porta}, G.~A. Voth, A.~M. Crawford, J.~Alexander, and E.~Bodenschatz.
\newblock Fluid particle acceleration in fully developed turbulence.
\newblock {\em Nature}, 409:\penalty0 1017, 2001.

\bibitem[Labb\'e et~al.(1996)Labb\'e, Pinton, and Fauve]{labbe96jphys}
R.~Labb\'e, J.-F. Pinton, and S.~Fauve.
\newblock Power fluctuations in turbulent swirling flows.
\newblock {\em J. Phys II}, 6:\penalty0 1099, 1996.

\bibitem[Lathrop et~al.(1992)Lathrop, Fineberg, and Swinney]{lathrop92}
D.~P. Lathrop, J.~Fineberg, and H.~L. Swinney.
\newblock Transition to shear-driven turbulence in {C}ouette-{T}aylor flow.
\newblock {\em Phys. Rev. A}, 46:\penalty0 6390, 1992.

\bibitem[Leprovost et~al.(2006)Leprovost, Dubrulle, and
  Chavanis]{leprovost2006}
N.~Leprovost, B.~Dubrulle, and P.-H. Chavanis.
\newblock Dynamics and thermodynamics of axisymmetric flows: Theory.
\newblock {\em Phys. Rev. E}, 73:\penalty0 046308, 2006.

\bibitem[Leprovost et~al.(2004)Leprovost, Mari\'e, and Dubrulle]{leprovost04}
N.~Leprovost, L.~Mari\'e, and B.~Dubrulle.
\newblock A stochastic model of torque in von {K}\'arm\'an swirling flow.
\newblock {\em Euro. Phys. J. B}, 39:\penalty0 121, 2004.

\bibitem[Lesieur(1990)]{lesieur}
M.~Lesieur.
\newblock {\em Turbulence in Fluids}.
\newblock Kluwer academic publishers, second revised edition edition, 1990.

\bibitem[Lohse and M\"uller-Groeling(1995)]{lohse1995}
D.~Lohse and A.~M\"uller-Groeling.
\newblock Bottleneck effects in turbulence: Scaling phenomena in $r$ versus $p$
  space.
\newblock {\em Phys. Rev. Lett.}, 74:\penalty0 1747, 1995.

\bibitem[Manneville(1990)]{manneville1990}
P.~Manneville.
\newblock {\em Dissipative structures and weak turbulence}.
\newblock Academic Press, 1990.

\bibitem[Mari{\'e}(2003)]{theselouis}
L.~Mari{\'e}.
\newblock {\em Transport de moment cin{\'e}tique et de champ magn{\'e}tique par
  un {\'e}coulement tourbillonaire turbulent: influence de la rotation}.
\newblock PhD thesis, Universit{\'e} Paris VII, 2003.

\bibitem[Mari{\'e} et~al.(2003)Mari{\'e}, Burguete, Daviaud, and
  L{\'e}orat]{marie2003epjb}
L.~Mari{\'e}, J.~Burguete, F.~Daviaud, and J.~L{\'e}orat.
\newblock Numerical study of homogeneous dynamo based on experimental von
  {K}\'arm\'an type flows.
\newblock {\em Euro. Phys. J. B}, 33:\penalty0 469, 2003.

\bibitem[Mari{\'e} and Daviaud(2004)]{marie04pof}
L.~Mari{\'e} and F.~Daviaud.
\newblock Experimental measurement of the scale-by-scale momentum transport
  budget in a turbulent shear flow.
\newblock {\em Phys. Fluids}, 16:\penalty0 457, 2004.

\bibitem[Mellor et~al.(1968)Mellor, Chapple, and Stokes]{mellor68}
G.~L. Mellor, P.~J. Chapple, and V.~K. Stokes.
\newblock On the flow between a rotating and a stationary disk.
\newblock {\em J. Fluid Mech.}, 31:\penalty0 95, 1968.

\bibitem[Moisy et~al.(2001)Moisy, Willaime, Andersen, and Tabeling]{moisy01}
F.~Moisy, H.~Willaime, J.~S. Andersen, and P.~Tabeling.
\newblock Passive scalar intermittency in low temperature helium flows.
\newblock {\em Phys. Rev. Lett.}, 86:\penalty0 4827, 2001.

\bibitem[Monchaux et~al.(2007)Monchaux, Berhanu, Bourgoin, Moulin, Odier,
  Pinton, Volk, Fauve, Mordant, P\'etr\'elis, Chiffaudel, Daviaud, Dubrulle,
  Gasquet, Mari\'e, and Ravelet]{monchaux2006b}
R.~Monchaux, M.~Berhanu, M.~Bourgoin, M.~Moulin, Ph. Odier, J.-F. Pinton,
  R.~Volk, S.~Fauve, N.~Mordant, F.~P\'etr\'elis, A.~Chiffaudel, F.~Daviaud,
  B.~Dubrulle, C.~Gasquet, L.~Mari\'e, and F.~Ravelet.
\newblock Generation of magnetic field by dynamo action in a turbulent flow of
  liquid sodium.
\newblock {\em Phys. Rev. Lett.}, 98:\penalty0 044502, 2007.

\bibitem[Monchaux et~al.(2006)Monchaux, Ravelet, Dubrulle, Chiffaudel, and
  Daviaud]{monchaux2006a}
R.~Monchaux, F.~Ravelet, B.~Dubrulle, A.~Chiffaudel, and F.~Daviaud.
\newblock Properties of steady states in turbulent axisymmetric flows.
\newblock {\em Phys. Rev. Lett.}, 96:\penalty0 124502, 2006.

\bibitem[Nikuradse(1932)]{nikuradse32}
J.~Nikuradse.
\newblock Gesetzmassigkeiten der turbulenten {S}tromungen in glatten {R}ohren.
\newblock {\em vDI Forshungsheft}, 356, 1932.
\newblock in English, in NASA TT F-10 (1966), p. 359.

\bibitem[Nikuradse(1933)]{nikuradse33}
J.~Nikuradse.
\newblock Stromungsgesetz in rauhren {R}ohren.
\newblock {\em vDI Forshungsheft}, 361, 1933.
\newblock in English, in Technical Memorandum 1292, National Advisory Committee
  for Aeronautics (1950).

\bibitem[Nore et~al.(2005)Nore, Moisy, and Quartier]{nore05}
C.~Nore, F.~Moisy, and L.~Quartier.
\newblock Experimental observation of near-heteroclinic cycles in the von
  {K}\'arm\'an swirling flow.
\newblock {\em Phys. Fluids}, 17:\penalty0 064103, 2005.

\bibitem[Nore et~al.(2004)Nore, Tartar, Daube, and Tuckerman]{nore04jfm}
C.~Nore, M.~Tartar, O.~Daube, and L.~S. Tuckerman.
\newblock Survey of instability thresholds of flow between exactly
  counter-rotating disks.
\newblock {\em J. Fluid Mech.}, 511:\penalty0 45, 2004.

\bibitem[Nore et~al.(2003)Nore, Tuckerman, Daube, and Xin]{nore03}
C.~Nore, L.~S. Tuckerman, O.~Daube, and S.~Xin.
\newblock The 1:2 mode interaction in exactly counter-rotating von {K}\'arm\'an
  swirling flow.
\newblock {\em J. Fluid Mech.}, 477:\penalty0 51, 2003.

\bibitem[Ouellette et~al.(2006)Ouellette, Xu, Bourgoin, and
  Bodenschatz]{ouellette2006}
N.~T. Ouellette, H.~Xu, M.~Bourgoin, and E.~Bodenschatz.
\newblock Small-scale anisotropy in {L}agrangian turbulence.
\newblock {\em New J. Phys.}, 8:\penalty0 102, 2006.

\bibitem[Pinton and Labb\'e(1994)]{pinton94}
J.-F. Pinton and R.~Labb\'e.
\newblock Correction to the {T}aylor hypothesis in swirling flows.
\newblock {\em J. Phys. II}, 4:\penalty0 1461, 1994.

\bibitem[Porter and Knobloch(2005)]{porter2005}
J.~Porter and E.~Knobloch.
\newblock Dynamics in the 1:2 spatial resonance with broken reflection
  symmetry.
\newblock {\em Physica D}, 201:\penalty0 318--344, 2005.

\bibitem[Ravelet(2005)]{theseflo}
F.~Ravelet.
\newblock {\em Bifurcations globales hydrodynamiques et
  magn\'etohydrodynamiques dans un \'ecoulement de von {K}\'arm\'an turbulent}.
\newblock PhD thesis, Ecole Polytechnique, France, 2005.

\bibitem[Ravelet et~al.(2005)Ravelet, Chiffaudel, Daviaud, and
  Leorat]{ravelet2005}
F.~Ravelet, A.~Chiffaudel, F.~Daviaud, and J.~Leorat.
\newblock Toward an experimental von {K}\'arm\'an dynamo: Numerical studies for
  an optimized design.
\newblock {\em Phys. Fluids}, 17:\penalty0 117104, 2005.

\bibitem[Ravelet et~al.(2004)Ravelet, Mari\'e, Chiffaudel, and
  Daviaud]{ravelet2004}
F.~Ravelet, L.~Mari\'e, A.~Chiffaudel, and F.~Daviaud.
\newblock Multistability and memory effect in a highly turbulent flow:
  Experimental evidence for a global bifurcation.
\newblock {\em Phys. Rev. Lett.}, 93:\penalty0 164501, 2004.

\bibitem[Ravelet et~al.(2007)Ravelet, Volk, Behranu, Chiffaudel, Daviaud,
  Dubrulle, Fauve, Monchaux, Mordant, Odier, P\'etr\'elis, and
  Pinton]{ravelet2007}
F.~Ravelet, R.~Volk, M.~Behranu, A.~Chiffaudel, F.~Daviaud, B.~Dubrulle,
  S.~Fauve, R.~Monchaux, N.~Mordant, Ph. Odier, F.~P\'etr\'elis, and J.-F.
  Pinton.
\newblock Magnetic induction in a turbulent flow of liquid sodium: mean
  behaviour and slow fluctuations.
\newblock submitted to Phys. Fluids, 2007.

\bibitem[Robert and Sommeria(1991)]{sommeria1991}
R.~Robert and J.~Sommeria.
\newblock Statistical equililbrium states for two-dimensional flows.
\newblock {\em J. Fluid Mech.}, 229:\penalty0 291--310, 1991.

\bibitem[Schlichting(1979)]{schlichting}
H.~Schlichting.
\newblock {\em Boundary-Layer Theory}.
\newblock McGraw-Hill, 7th ed. edition, 1979.

\bibitem[Schouveiler et~al.(2001)Schouveiler, Gal, and Chauve]{schouveiler01}
L.~Schouveiler, P.~Le Gal, and M.-P. Chauve.
\newblock Instabilities of the flow between a rotating and a stationary disk.
\newblock {\em J. Fluid Mech.}, 443:\penalty0 329, 2001.

\bibitem[S{\o}rensen and Christensen(1995)]{sorensen95}
J.~B. S{\o}rensen and E.~A. Christensen.
\newblock Direct numerical simulation of rotating fluid flow in a closed
  cylinder.
\newblock {\em Phys. Fluids}, 7:\penalty0 764, 1995.

\bibitem[Spohn et~al.(1998)Spohn, Mory, and Hopfinger]{spohn98}
A.~Spohn, M.~Mory, and E.~J Hopfinger.
\newblock Experiments on vortex breakdown in a confined flow generated by a
  rotating disc.
\newblock {\em J. Fluid Mech.}, 370:\penalty0 73, 1998.

\bibitem[Stewartson(1953)]{stewartson53}
K.~Stewartson.
\newblock On the flow between two rotating coaxial disks.
\newblock {\em Proc. Camb. Phil. Soc.}, 49:\penalty0 333, 1953.

\bibitem[Tabeling et~al.(1996)Tabeling, Zocchi, Belin, Maurer, and
  Willaime]{tabeling96}
P.~Tabeling, G.~Zocchi, F.~Belin, J.~Maurer, and H.~Willaime.
\newblock Probability density functions{,} skewness and flatness in large
  {R}eynolds number turbulence.
\newblock {\em Phys. Rev. E}, 53:\penalty0 1613, 1996.

\bibitem[Tennekes and Lumley(1972)]{tennekes}
H.~Tennekes and J.~L. Lumley.
\newblock {\em A first course in turbulence}.
\newblock MIT Press, 1972.

\bibitem[Titon and Cadot(2003)]{titon03pof}
J.-H. Titon and O.~Cadot.
\newblock The statistics of power injected in a closed turbulent flow: Constant
  torque forcing versus constant velocity forcing.
\newblock {\em Phys. Fluids}, 15:\penalty0 625, 2003.

\bibitem[Volk et~al.(2006)Volk, Odier, and Pinton]{volk2006pof}
R.~Volk, P.~Odier, and J.-F. Pinton.
\newblock Fluctuation of magnetic induction in von {K}\'arm\'an swirling flows.
\newblock {\em Phys. Fluids}, 18:\penalty0 085105, 2006.

\bibitem[von K{\'a}rm{\'a}n(1921)]{vonkarman21}
T.~von K{\'a}rm{\'a}n.
\newblock {\"U}ber lamin{\"a}re und turbulente {R}eibung.
\newblock {\em Z. Angew. Math. Mech.}, 1:\penalty0 233, 1921.

\bibitem[Zandbergen and Dijkstra(1987)]{zandbergen87}
P.~J. Zandbergen and D.~Dijkstra.
\newblock von {K}\'arm\'an swirling flows.
\newblock {\em Annu. Rev. Fluid Mech.}, 19:\penalty0 465, 1987.

\bibitem[Zocchi et~al.(1994)Zocchi, Tabeling, Maurer, and Willaime]{zocchi94}
G.~Zocchi, P.~Tabeling, J.~Maurer, and H.~Willaime.
\newblock Measurement of the scaling of the dissipation at high {R}eynolds
  numbers.
\newblock {\em Phys. Rev. E}, 50:\penalty0 3693, 1994.

\end{thebibliography}

\end{document}